\RequirePackage{lineno}
\documentclass[aps,twocolumn,superscriptaddress,preprintnumbers,amsmath,amssymb]{revtex4}
\usepackage{graphicx}
\usepackage{epsfig}
\usepackage{dcolumn}
\usepackage{bm}
\usepackage{amsmath}
\usepackage{color}
\usepackage[colorlinks,linkcolor=red,anchorcolor=green,citecolor=blue]{hyperref}
\usepackage{multirow}
\usepackage{relsize}
\usepackage{tabularx}
\usepackage{lipsum}
\usepackage{ulem}
\usepackage{orcidlink}
\newcommand{\lum}{{\cal L}}
\newcommand{\eff}{\varepsilon}
\newcommand{\BF}{{\cal B}}

\newcommand{\fours}{\Upsilon(4S)}

\newcommand{\EE}{e^+e^-}
\newcommand{\MM}{\mu^+\mu^-}
\newcommand{\LL}{\ell^+\ell^-}
\newcommand{\pipi}{\pi^+\pi^-}
\newcommand{\kk}{K^+K^-}
\newcommand{\pp}{p\bar{p}}
\newcommand{\hh}{h^+h^-}
\newcommand{\jpsi}{J/\psi}
\newcommand{\psip}{\psi(2S)}
\newcommand{\pipijpsi}{\pi^+\pi^-J/\psi}
\newcommand{\kkjpsi}{K^+K^-J/\psi}
\newcommand{\ppjpsi}{p\bar{p}J/\psi}
\newcommand{\hhjpsi}{h^+h^-J/\psi}
\newcommand{\hhjpsix}{h^+h^-J/\psi~(h=\pi/K/p)}

\newcommand{\jpsill}{J/\psi\to\LL}
\newcommand{\infb}{\rm fb^{-1}}

\newcommand{\gev}{\rm GeV}

\newcommand{\gevcs}{{\rm GeV}/c^2}

\newcommand{\LK}{\mathcal{L}}
\newcommand{\ecms}{\sqrt{s}}

\begin{document}

\title{\quad\\[0.5cm]
Study of $e^{+}e^{-}\to h^{+}h^{-}J/\psi~(h=\pi,~K,~p)$ via initial-state radiation at Belle~II
}

\author{M.~Abumusabh\,\orcidlink{0009-0004-1031-5425}} % 26883
  \author{I.~Adachi\,\orcidlink{0000-0003-2287-0173}} % 2590
% \author{K.~Adamczyk\,\orcidlink{0000-0001-6208-0876}} % 2239
  \author{A.~Aggarwal\,\orcidlink{0000-0002-5623-3896}} % 24463
  \author{L.~Aggarwal\,\orcidlink{0000-0002-0909-7537}} % 10083
% \author{P.~Ahlburg\,\orcidlink{0000-0002-9832-7604}} % 2367
  \author{H.~Ahmed\,\orcidlink{0000-0003-3976-7498}} % 11323
% \author{J.~K.~Ahn\,\orcidlink{0000-0002-5795-2243}} % 7423
  \author{Y.~Ahn\,\orcidlink{0000-0001-6820-0576}} % 14363
  \author{H.~Aihara\,\orcidlink{0000-0002-1907-5964}} % 2223
  \author{N.~Akopov\,\orcidlink{0000-0002-4425-2096}} % 9443
  \author{S.~Alghamdi\,\orcidlink{0000-0001-7609-112X}} % 27804
  \author{M.~Alhakami\,\orcidlink{0000-0002-2234-8628}} % 28103
% ...
% 13403
% \author{S.~Zou\,\orcidlink{0000-0003-3377-7222}} % 19363
\collaboration{The Belle II Collaboration}

\begin{abstract}

%\linenumbers
Using a data sample of 427.9 fb$^{-1}$ collected by the Belle~II detector at or near the $\Upsilon(4S)$ and $\Upsilon(10753)$ resonances, the cross sections for $e^+e^-\to h^+h^-J/\psi$ $(h=\pi/K/p)$ at center-of-mass energies ranging from 3.8 GeV or the production threshold to 5.5/6.0/7.0 GeV have been measured via initial-state radiation. The cross sections for the processes $e^+e^-\to \pi^+\pi^-J/\psi$ and $e^+e^-\to K^+K^-J/\psi$ are consistent with previously published results. The cross sections for these channels obtained by combining with previous Belle results are also given. The process $e^+e^-\to p\bar p J/\psi$ is investigated for the first time. The yields are small and no significant structure is observed in the cross section versus energy. Searches for vector charmonium-like states in the $\hhjpsi$ systems, and for associated intermediate states in the $h^{\pm}\jpsi$ systems, are also presented.

\end{abstract}

\maketitle

%\linenumbers

% ---------------------
%     Introdution
% ---------------------
\section{Introduction}
The traditional quark model successfully describes most charmonium states above the open-charm threshold ($D\bar{D}$ production threshold), such as $\psi(3770)$, $\psi(4040)$, and $\psi(4160)$~\cite{Gell-Mann:1964,Fritzsch:1973}. However, since 2003, many unexpected resonant states have been observed. These states exhibit exotic properties, including unusual masses and quantum numbers, which cannot be explained by the simple $c\bar{c}$ configuration within the traditional quark model~\cite{Brambilla:2019,Chen:2023}. Among these, several vector charmonium-like states have been identified through either direct $\EE$ annihilation or initial-state radiation (ISR) processes, in which the $\EE$ system emits one or more photons, resulting in a lower center-of-mass (c.m.)\ energy for the annihilating $\EE$ pair~\cite{BaBar:2005,Yuan:2007,Wang:2007,Pakhlova:2010,Liu:2013,Wang:2015,Jia:2019,Jia:2020}. Candidates for tetraquark and pentaquark states have also been discovered~\cite{Liu:2013,Choi:2008,Mizuk:2009,Ablikim:2013,Ablikim:2017,Ablikim:2021,Aaij:2015,Aaij:2019,Aaij:2021,Aaij:2022,Aaij:2023}. These states have unusual decay modes and are close to production thresholds for two-hadron systems.  A variety of phenomenological models have been proposed to describe their structures, including the compact multi-quark model~\cite{Esposito:2016} and the hadronic molecule model~\cite{Guo:2017,Chen:2022}. Notably, the first observed vector charmonium-like state, $Y(4260)$, also known as $\psi(4230)$~\cite{PDG:2024}, has been interpreted as a $D_1(2420)\bar{D}$ molecular state~\cite{Wang:2013}, and it has been suggested that the pentaquark candidates $P_{c}^{+}$ are molecular states composed of $\Lambda_c^{(*)}\bar{D}^{(*)}$ or $\Sigma_c^{(*)}\bar{D}^{(*)}$~\cite{Wu:2010,Wu:2011,Liu:2016}. However, due to the limited availability of experimental data, a comprehensive understanding of these exotic states remains elusive, and the validity of the proposed models continues to be a subject of active debate~\cite{Brambilla:2019,PDG:2024}.

The $Y(4260)$ state was first discovered in the $\pipijpsi$ final state by BaBar in 2005~\cite{BaBar:2005}. Later, this state was found by BESIII to consist of two sub-structures, $Y(4230)$ and $Y(4320)$~\cite{BESIII:2017,BESIII:2022}. In addition to these findings, both Belle and BESIII observed a broad resonance in the $\pipijpsi$ system, namely $Y(4008)$~\cite{Yuan:2007,Liu:2013,BESIII:2022}. However, BaBar disputed the existence of this state~\cite{BaBar:2012}. A global coupled-channel analysis predicts a $\psi(4040)$-like enhancement in the $\pipijpsi$ invariant mass distribution~\cite{Nakamura:2023obk}. Investigations of related strange-charmonium final states have been carried out through the processes $\EE\to\kkjpsi$ and $\EE\to K^0_SK^0_S\jpsi$ at BESIII. These studies have led to the observation of two novel structures, $Y(4500)$ and $Y(4710)$~\cite{KKJpsi:2018,KKJpsi:2022,KsKsJpsi:2023,KKJpsi:2023}. Belle also investigated the processes $\EE\to\kkjpsi$ and $\EE\to K^0_SK^0_S\jpsi$ via ISR, but did not have sufficient sensitivity to confirm these structures~\cite{Yuan:2008,Shen:2014}. 

Additional measurements of the $\hhjpsi$ cross sections are essential input for the investigation of these exotic states. The $\EE\to\pipijpsi$ cross section was measured by BaBar from 3.5 to 5.5~GeV and by Belle from 3.8 to 5.5~GeV using ISR production in data samples with integrated luminosities of $454~\mathrm{fb}^{-1}$~\cite{BaBar:2012} and $967~\mathrm{fb}^{-1}$~\cite{Liu:2013}, respectively. BESIII measured it from 3.77 to 4.70~GeV using an energy-scan method~\cite{BESIII:2017,BESIII:2022}. For the $\EE\to\kkjpsi$ cross section, Belle measured it from the production threshold up to 6.0~GeV with a data sample of $980~\mathrm{fb}^{-1}$~\cite{Shen:2014}, and BESIII measured it from the production threshold to 4.95~GeV using the energy-scan method~\cite{KKJpsi:2022,KKJpsi:2023}. Recently, the authors of Ref.~\cite{Zhang:2025pfz} predicted the cross section of $\EE\to\ppjpsi$ to be ${\cal O}(4~\mathrm{fb})$ around 6~GeV. Independent experimental measurements from Belle~II can add to and extend these existing measurements.

We present a study of the process $\EE\to\hhjpsix$ via ISR by analyzing $427.9~\infb$ of data collected with the Belle~II detector at or near the $\fours$ and $\Upsilon(10753)$ resonances~\cite{luminosity:2025}. The analysis focuses on c.m.\ energies ranging from 3.8 GeV or the production threshold up to 5.5/6.0/7.0 GeV. We measure the production cross sections of $\EE\to\hhjpsi$ processes in the regions of several vector charmonium-like states. This analysis provides an independent cross check of the processes $\EE\to\pipijpsi$ and $\EE\to\kkjpsi$, with precision comparable to that of Belle. We also investigate the previously unexplored final state involving baryons, $\EE\to\ppjpsi$, which may offer deeper insight into pentaquark production mechanisms. Furthermore, we explore possible intermediate states in the $h^{\pm}\jpsi$ systems.

% -----------------------------
%     Detector and Dataset
% -----------------------------
\section{Detector and dataset}
The Belle~II experiment is located at SuperKEKB, which collides electrons and positrons at and near the $\Upsilon(4S)$ resonance~\cite{SuperKEKB}. The Belle~II detector~\cite{BelleII} has a cylindrical geometry and includes a two-layer silicon-pixel detector~(PXD) surrounded by a four-layer double-sided silicon-strip detector~(SVD)~\cite{Belle-IISVD:2022upf} and a 56-layer central drift chamber~(CDC). Position information from these detectors is used to reconstruct the trajectories of charged particles~\cite{Bertacchi:2020eez}. Only one sixth of the second layer of the PXD was installed for the data analyzed here. The symmetry axis of these detectors, defined as the $z$ axis, is almost coincident with the direction of the electron beam. Surrounding the CDC, which also provides ${\rm d}E/{\rm d}x$ energy-loss measurements, is a time-of-propagation counter~(TOP)~\cite{Kotchetkov:2018qzw} in the central region and an aerogel-based ring-imaging Cherenkov counter~(ARICH) in the forward region. These detectors provide charged-particle identification.  Surrounding the TOP and ARICH is an electromagnetic calorimeter~(ECL) based on CsI(Tl) crystals that primarily provides energy and timing measurements for photons and electrons. Outside of the ECL is a superconducting solenoid magnet. The magnet provides a 1.5~T magnetic field that is parallel to the $z$ axis. Its flux return is instrumented with resistive-plate chambers and plastic scintillator modules to detect muons, $K^0_L$ mesons, and neutrons.

Signal Monte Carlo (MC) samples are generated using a combination of {\sc phokhara}~\cite{phokhara:2002,Campanario:2019mjh} and {\sc EvtGen}~\cite{evtgen:2001}, and are employed to guide event selection and determine detection efficiencies. The simulation of ISR events is performed with {\sc phokhara} at next-to-leading order accuracy in quantum electrodynamics, while meson and hadron decays are generated using {\sc EvtGen}. Effects of final-state radiation (FSR) are incorporated with {\sc Photos}~\cite{photos:1991}. The {\sc geant4}~\cite{geant4:2003} package is employed to simulate the passage of particles through the detector and the detector response. One million signal events per channel are generated for each of $\pi^{+}\pi^{-}J/\psi$, $K^{+}K^{-}J/\psi$, and $p\bar{p}J/\psi$ at various center-of-mass energies ($\sqrt{s}$), with the $\hhjpsi$ system distributed according to phase space (PHSP). For each channel, events are produced at eight different energy points to determine the detection efficiency. The dependence of efficiency on $\sqrt{s}$ is determined from a fit to the eight discrete values. These efficiencies are further weighted by the Dalitz plot distributions obtained from data to derive the final efficiency estimates, which will be described in detail later in the text. Generic MC samples including background events from $\EE\to q\bar{q}$~($q=u,~d,~s,~c$), $\EE\to\tau^+\tau^-$, $\EE\to(\gamma)\EE$ (Bhabha), $\EE\to(\gamma)\MM$ (di-muon), and $\EE\to\hh$ $(h=\pi,~K,~p)$ via ISR, with the same integrated luminosity as the data, are utilized to study background processes in the analysis~\cite{Sjostrand:2014zea,Jadach:1999vf,Jadach:1990mz}. All data and simulated events are reconstructed and analyzed using the Belle~II Analysis Software Framework~\cite{basf2:2019}.

% -----------------------
%     Event Selection
% -----------------------
\section{Event selection}
Events are selected by the hardware-based level~1 triggers using signals from the CDC, ECL, or KLM. Tracks are required to be within $|d_z|<3~{\rm cm}$ and $|d_r|<1~{\rm cm}$, where $d_z$ and $d_r$ are the distances of the closest approach to the interaction point along the $z$-axis and in the transverse plane, respectively. Tracks are also required to have a polar angle ($\theta$) between $17^\circ$ and $150^\circ$, which is within the acceptance of the CDC. We require the total number of selected tracks in the event to be four, and the net charge to be zero. The particle identification (PID) information from various detector subsystems is combined in a likelihood $\LK_{i}$ for particle species $i\in\{e,~\mu,~\pi,~K,~p,~d\}$~\cite{Belle-II:2025tpe}. Tracks with ${\cal R}_{\pi}=\LK_\pi / ( \LK_\pi + \LK_K )>0.6~(<0.4)$ are identified as pions (kaons). Pions (kaons) are identified with approximately 95\% (92\%) efficiency and approximately 6\% (4\%) kaon (pion) mis-identification efficiency. Tracks with ${\cal R}_{e/\mu/p} = \LK_{e/\mu/p} / \sum^6_i\LK_i >0.5$ are identified as electrons, muons, and protons, respectively, corresponding to approximate identification efficiencies of 90\%, 85\%, and 90\%, respectively. For electrons and positrons, bremsstrahlung and FSR corrections are applied by incorporating the four momenta of the photons within a cone of 50 mrad around the initial direction of the electron or positron. Photons are identified from ECL energy deposits greater than 75, 100 and 50 MeV in the forward, backward and barrel regions, respectively. These regions correspond to $\theta$ ranges in the lab frame of $[12.4^\circ,~31.4^\circ]$, $[130.7^\circ,~155.1^\circ]$, and $[32.2^\circ,~128.7^\circ]$.

The $\jpsi$ candidates are reconstructed in both $\EE$ and $\MM$ decay modes. A clear $\jpsi$ signal is observed in the data, with a purity exceeding 95\% for the process $\psip\to\pipijpsi$, and approximately 80\% for the process $\EE\to\pipijpsi$ in the high-mass region. The $\jpsi$ signal regions for the two modes are defined to have $M(\EE)$ in the region $(3.02,~3.14)~\gevcs$ and $M(\MM)$ in the region $(3.06,~3.14)~\gevcs$. The $\jpsi$ mass sidebands are defined as having $M(\EE)$ in either the $(2.81,~2.99)~\gevcs$ or the $(3.17,~3.35)~\gevcs$ regions, or having $M(\MM)$ in either the $(2.91,~3.03)~\gevcs$ or the $(3.17,~3.29)~\gevcs$ regions. The width of each sideband region is three times the width of the corresponding signal region. A mass-constrained fit to $\jpsi$ candidates is performed to determine $\jpsi$-related variables and improve the resolution.

In the channel $\EE\to\pipi\jpsi
~(\to\EE)$, background from the process $\EE\to\EE\gamma$, in which photons convert to $\EE$ pairs via interaction with the detector material and the resulting pairs are misidentified as $\pipi$, is suppressed by requiring ${\cal R}_{e}<0.5$ and $\LK_{e}<\LK_{\pi}$ for pions. These conversion background events are further reduced by requiring the invariant mass of $\pipi$ to be larger than 0.35 $\gevcs$. In $\EE\to\kk\jpsi~(\to\EE)$, the invariant mass of $\kk$ is required to be larger than 1.05 $\gevcs$ to veto $\gamma$ conversion events and remove the events with a $\phi~(\to\kk)$ meson in the final state. The ISR events are identified by the requirement $|M^2_{\rm recoil}|<1~(\gevcs)^2$, where $M^2_{\rm recoil}=(p_{\rm c.m.}-p_{h^+}-p_{h^-}-p_{\ell^+}-p_{\ell^-})^2$. Here, $p_i$ represents the four-momentum of the corresponding particle in the $\EE$ c.m.\ frame, and $p_{\rm c.m.}$ is the four-momentum of the initial $\EE$ system. After event selection, the fraction of events with multiple candidates is zero in both the signal MC and data samples.

\begin{figure}[htpb]
    \centering
    \includegraphics[width=7.0cm]{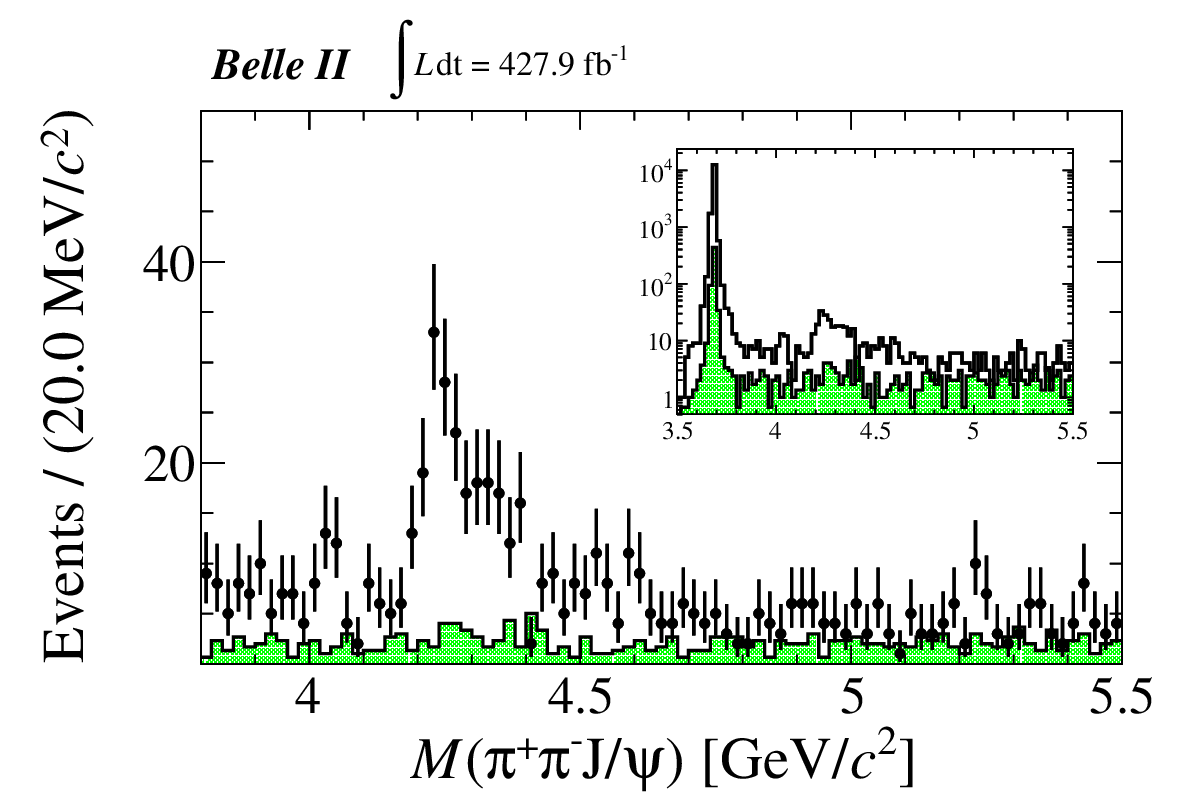}
    \put(-150, 100){\bf (a)} \\
    \includegraphics[width=7.0cm]{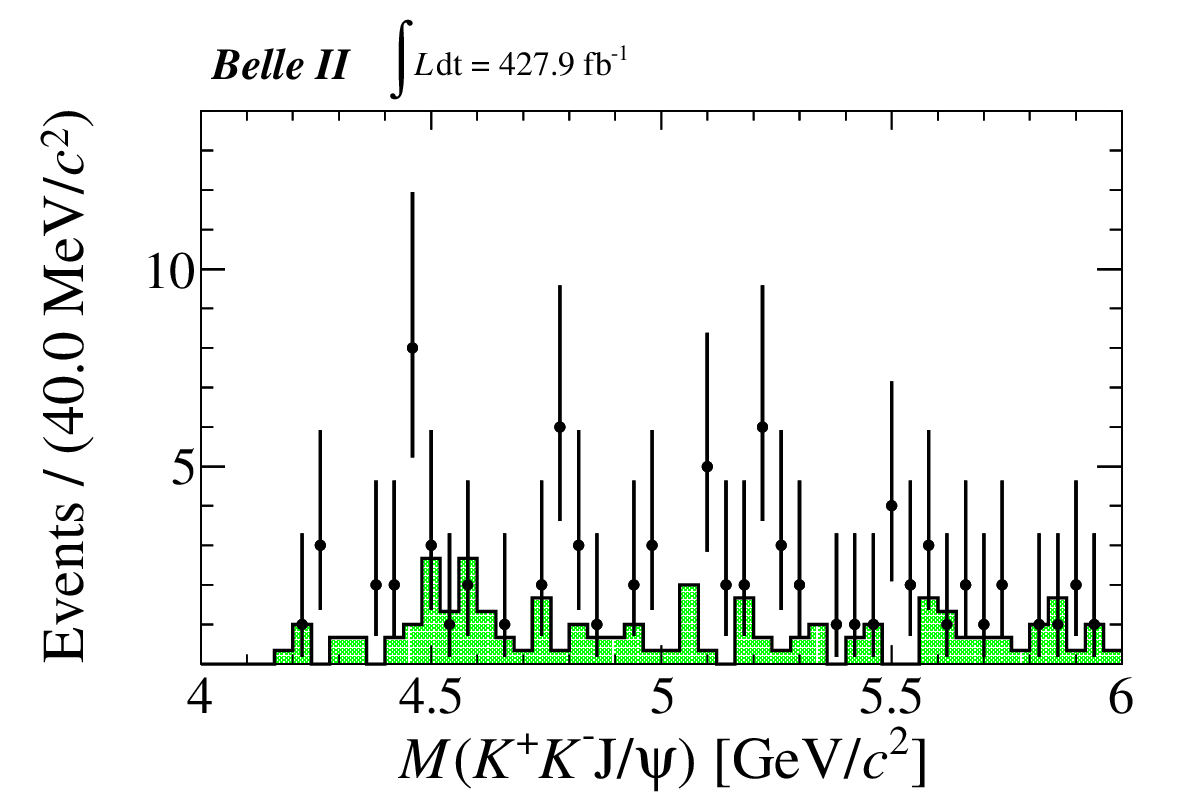}
    \put(-150, 100){\bf (b)} \\
    \includegraphics[width=7.0cm]{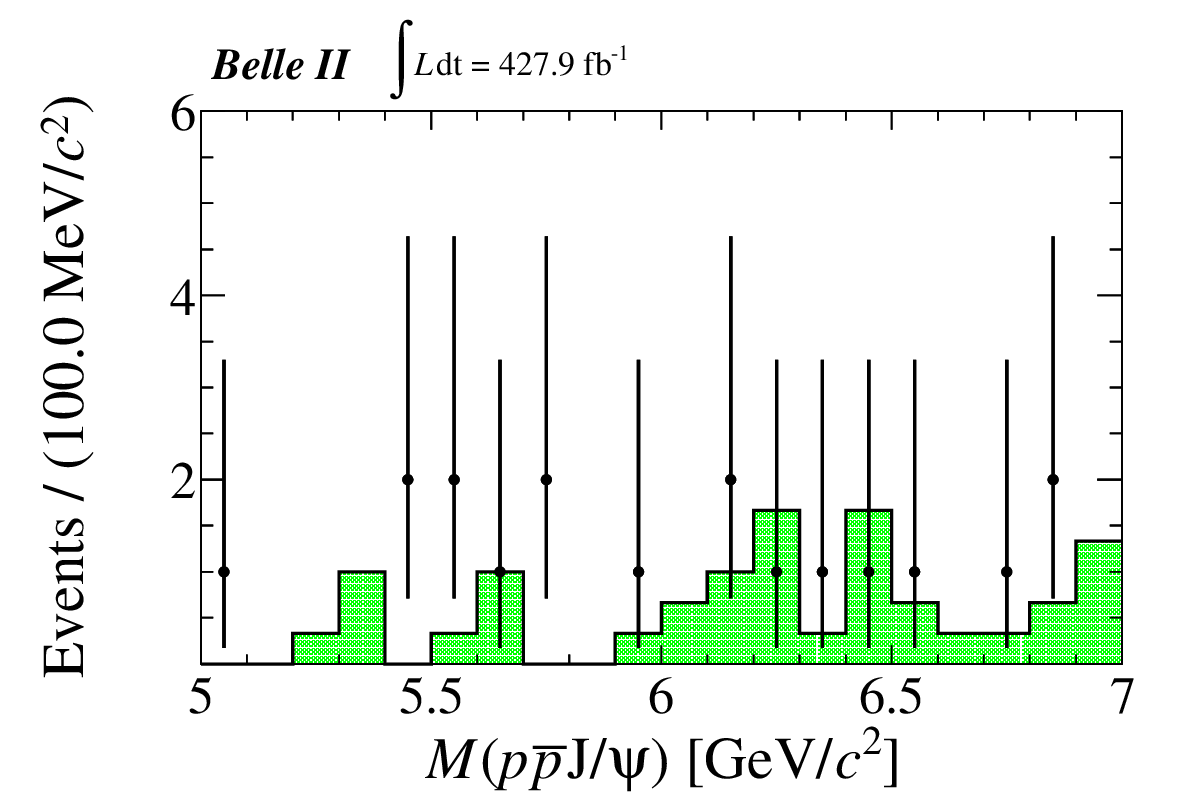}
    \put(-150, 100){\bf (c)}
    \caption{Invariant mass distributions of (a) $\pipijpsi$, (b) $\kkjpsi$, and (c) $\ppjpsi$. The black dots with error bars represent the selected data and the green shaded histograms are from the normalized $J/\psi$ sidebands. The inset in (a) shows the distribution extended to lower masses on a logarithmic vertical scale, and the large peak near 3.686 GeV is from $\gamma_{\rm ISR}\psi(2S)$.}
    \label{fig:mhhll}
\end{figure}

After applying all selection criteria, the $\hhjpsi$ invariant mass distributions for the $\jpsi$ candidates within the signal window, together with the backgrounds estimated from the normalized $\jpsi$ mass sidebands, are shown in Fig.~\ref{fig:mhhll}. These background events are employed to estimate the combinatorial background arising from non-$\jpsi$ contributions. The background contamination due to cross-feed among the three signal channels has been examined and found to be negligible using MC simulated events. The remaining background comes mostly from processes that occur through two virtual photons such as $\EE\to\gamma\gamma^*\gamma^*\to\gamma VV$, for example $\EE\to\gamma\rho\jpsi$ and $\EE\to\gamma\phi\jpsi$; however, these contributions have been calculated to be small~\cite{Zhu:2008}. We investigate them using MC simulations, confirming that their contributions are insignificant and can be neglected in this study. The events with a real $\jpsi$ and a broad pseudo-scalar resonance are considered as signal. 

\begin{figure}[htpb]
    \centering
    \includegraphics[width=7.5cm]{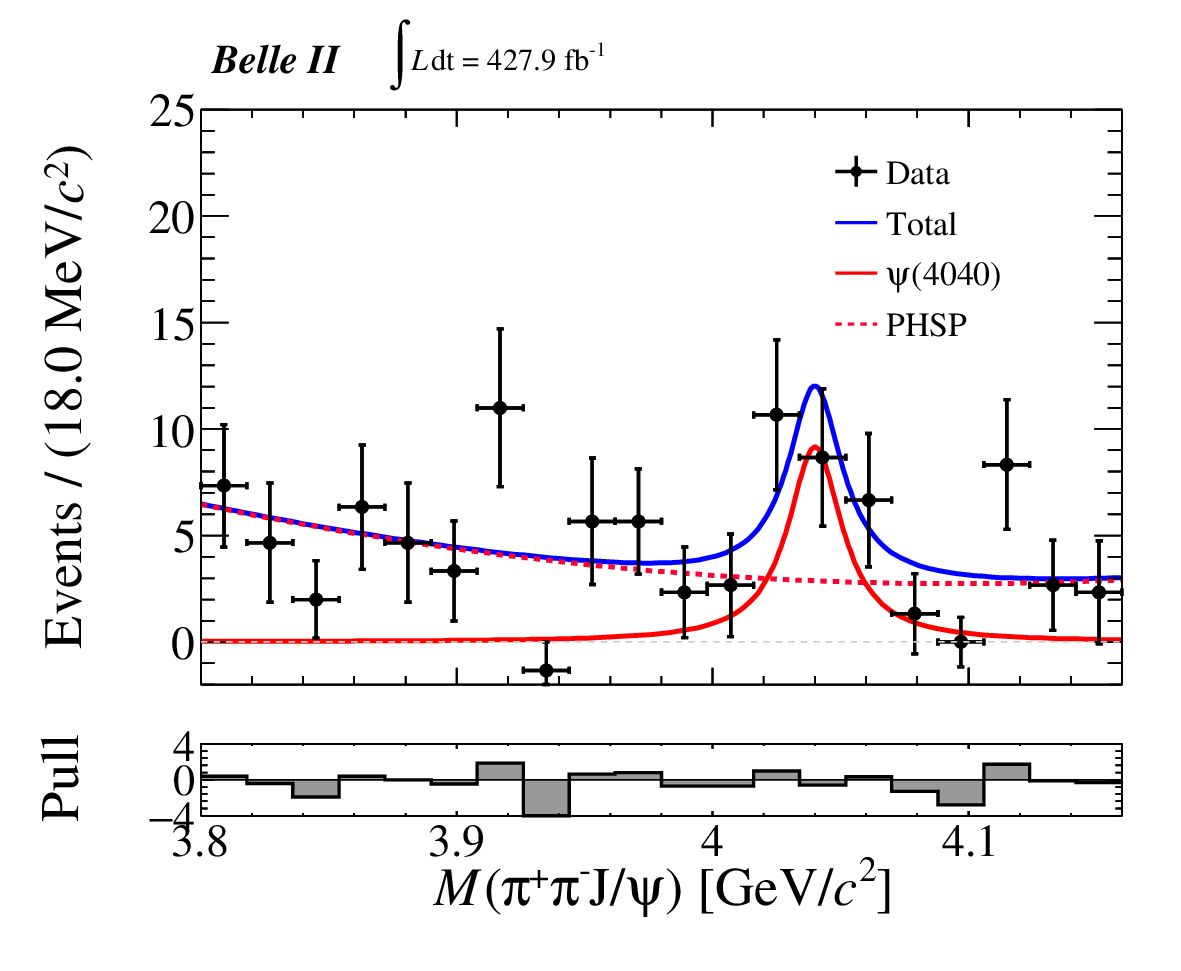}
    \caption{An unbinned extended maximum likelihood fit to the invariant mass distribution of $\pipijpsi$ with the signal of $\psi(4040)$ state. The black dots with error bars are the $\pipijpsi$ signal candidates, obtained by subtracting the background estimated from the normalized $\jpsi$ mass sidebands.}
    \label{fig:psi4040}
\end{figure}

We observe an enhancement around 4.26 GeV in the $\pipijpsi$ invariant mass distribution, consistent with the $Y(4230/4320)$ structure previously reported by Belle~\cite{Yuan:2007,Liu:2013} and BESIII~\cite{BESIII:2017,BESIII:2022}. In addition, the observed signal events around 4.1 GeV show an excess over the expected background, which is consistent with the predicted $\psi(4040)$-like enhancement~\cite{Nakamura:2023obk}. To investigate the presence of the $\psi(4040)$, we perform an unbinned extended maximum likelihood fit, as shown in Fig.~\ref{fig:psi4040}. The fit range is restricted to the invariant mass below 4.16 GeV$/c^2$ in order to suppress the effects of higher-mass resonances. The $\psi(4040)$ signal is modeled with a Breit-Wigner function, whose parameters are fixed to the world average values~\cite{PDG:2024}. The background from non-$\jpsi$ events is estimated using the normalized $\jpsi$ mass sideband regions. A second-order Chebyshev polynomial is used to describe the PHSP $\pipijpsi$ production. The statistical significance of the $\psi(4040)$ state is found to be $2.0\sigma$. For the $\kkjpsi$ process, no significant structure is observed in the invariant mass distribution. No clear signal of $\ppjpsi$ production is found.

% -------------------------------
%    Intermediate State Search
% -------------------------------
\section{Intermediate state search}
\begin{figure}[htpb]
    \centering
    \includegraphics[width=6.5cm]{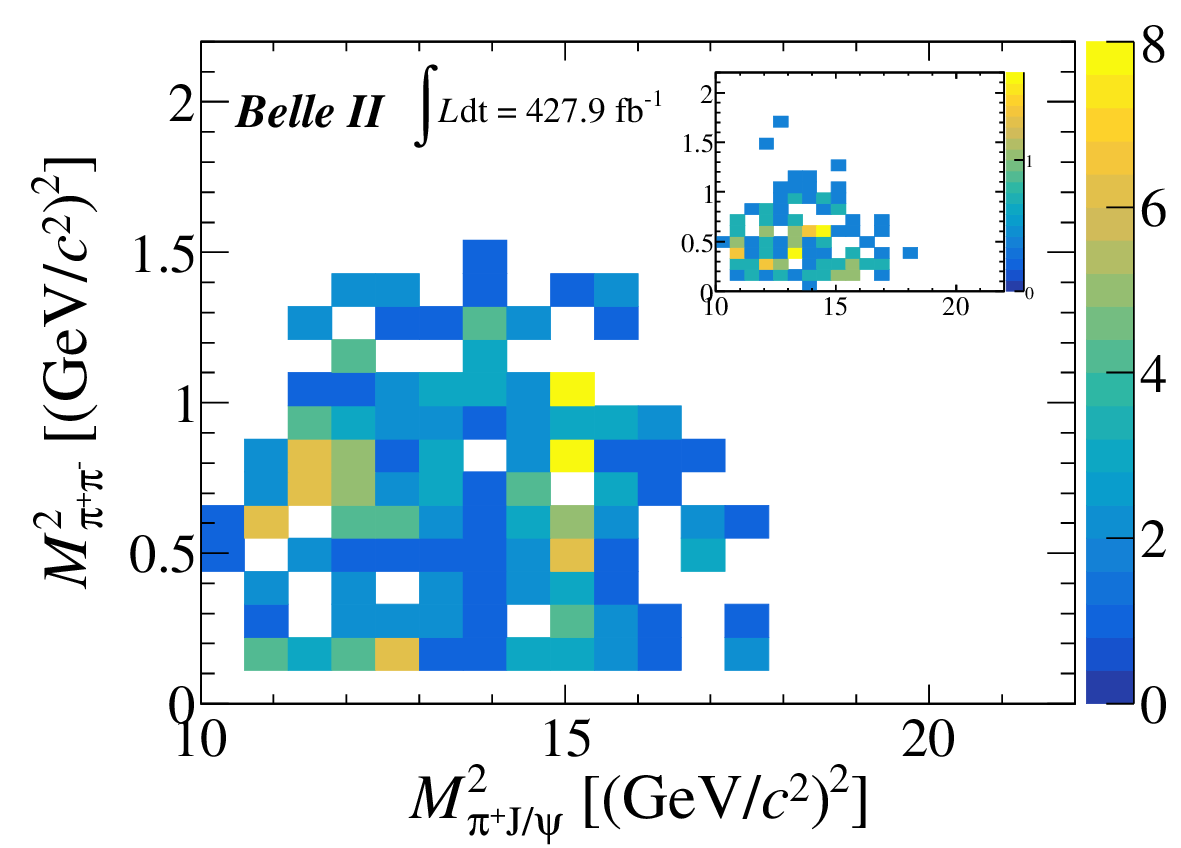}
    \put(-40, 35){\bf (a)} \\
    \includegraphics[width=6.5cm]{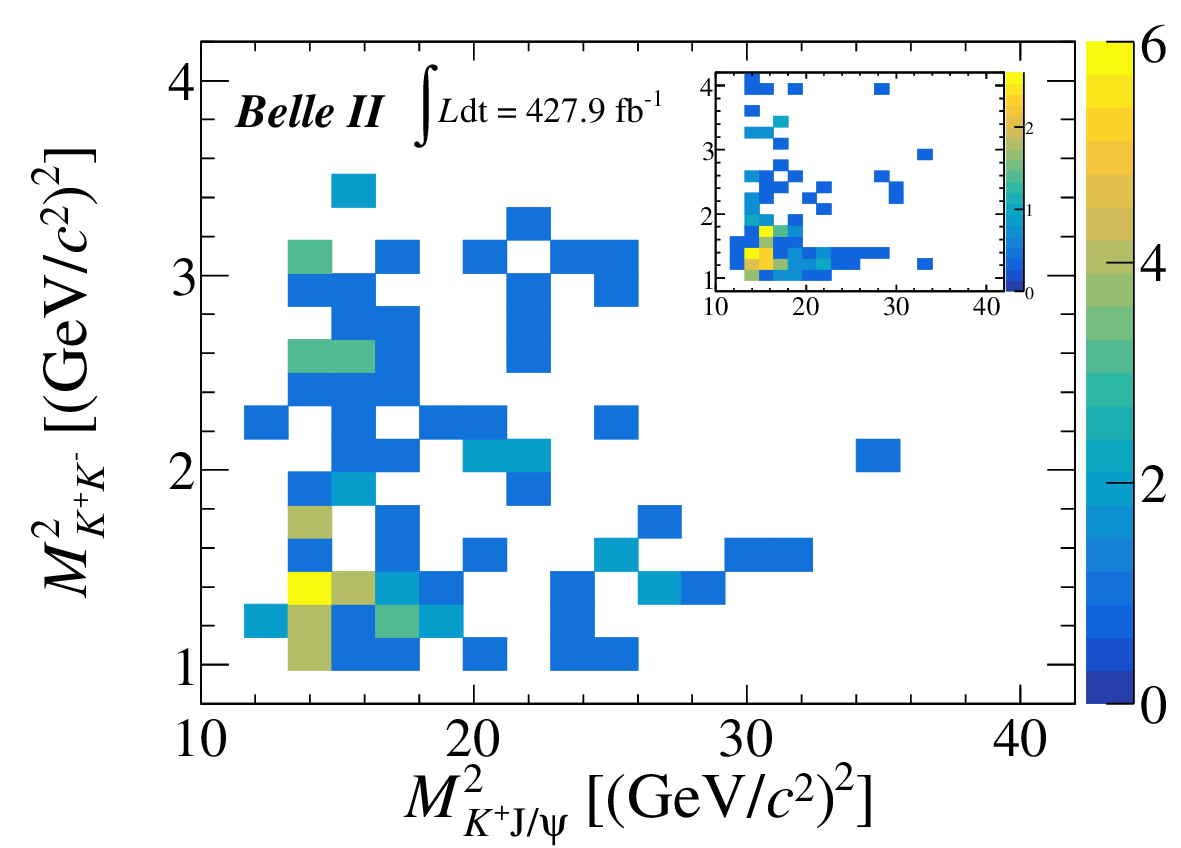}
    \put(-40, 35){\bf (b)}
    \caption{The Dalitz plots of (a) $\EE\to\pipijpsi$ in the $Y(4230/4320)$ region and (b) $\EE\to\kkjpsi$ with $M(\kkjpsi)<6.0~\gevcs$. The insets show the background events from the normalized $\jpsi$ mass sidebands.}
    \label{fig:dalitz}
\end{figure}

\begin{figure}[htpb]
    \centering
    \includegraphics[width=7.0cm]{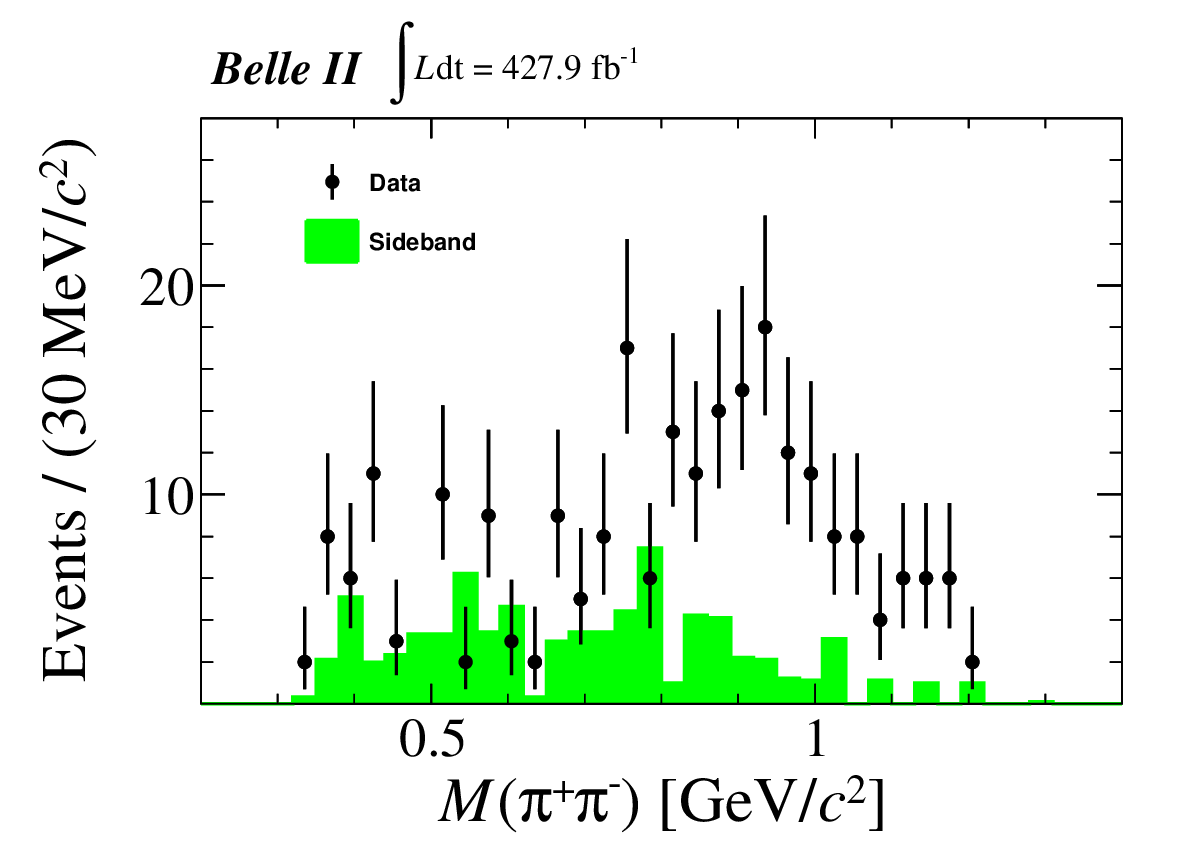}
    \caption{The distribution of the $M(\pipi)$ in the $Y(4230/4320)$ region.}
    \label{fig:mpipi}
\end{figure}

The presence of potential intermediate states is investigated by analyzing the projections of the Dalitz plots~\cite{Dalitz:1953cp} for the selected $\hhjpsi$ candidate events. Figure~\ref{fig:dalitz} presents the Dalitz plots in the $\jpsi$ signal and sideband regions. In total 235 (43) events are obtained in the $\jpsi$ signal (sideband) region for the process $\EE\to\pipijpsi$ in the $Y(4230/4320)$ region, defined as having $M(\pipijpsi)$ in the region $(4.15,~4.45)~\gevcs$, and 96 (43) events are obtained in the $\jpsi$ signal (sideband) regions for the process $\EE\to\kkjpsi$ with $M(\kkjpsi)<6.0~\gevcs$. For the $\EE\to\pipijpsi$ events in the $Y(4260)$ signal region, the $\pipi$ invariant mass distribution shown in Fig.~\ref{fig:mpipi} is similar to what was observed by Belle~\cite{Yuan:2007,Liu:2013}, where the contribution from the process $\EE\to\jpsi f_0(980)~(\to\pipi)$ can be seen. In addition, band-like excesses around $M_{\jpsi\pi} = 3.9~{\rm GeV}/c^2$ are observed in the $\pi^{\pm}\jpsi$ system.

\begin{figure}[htpb]
    \centering
    \includegraphics[width=7.0cm]{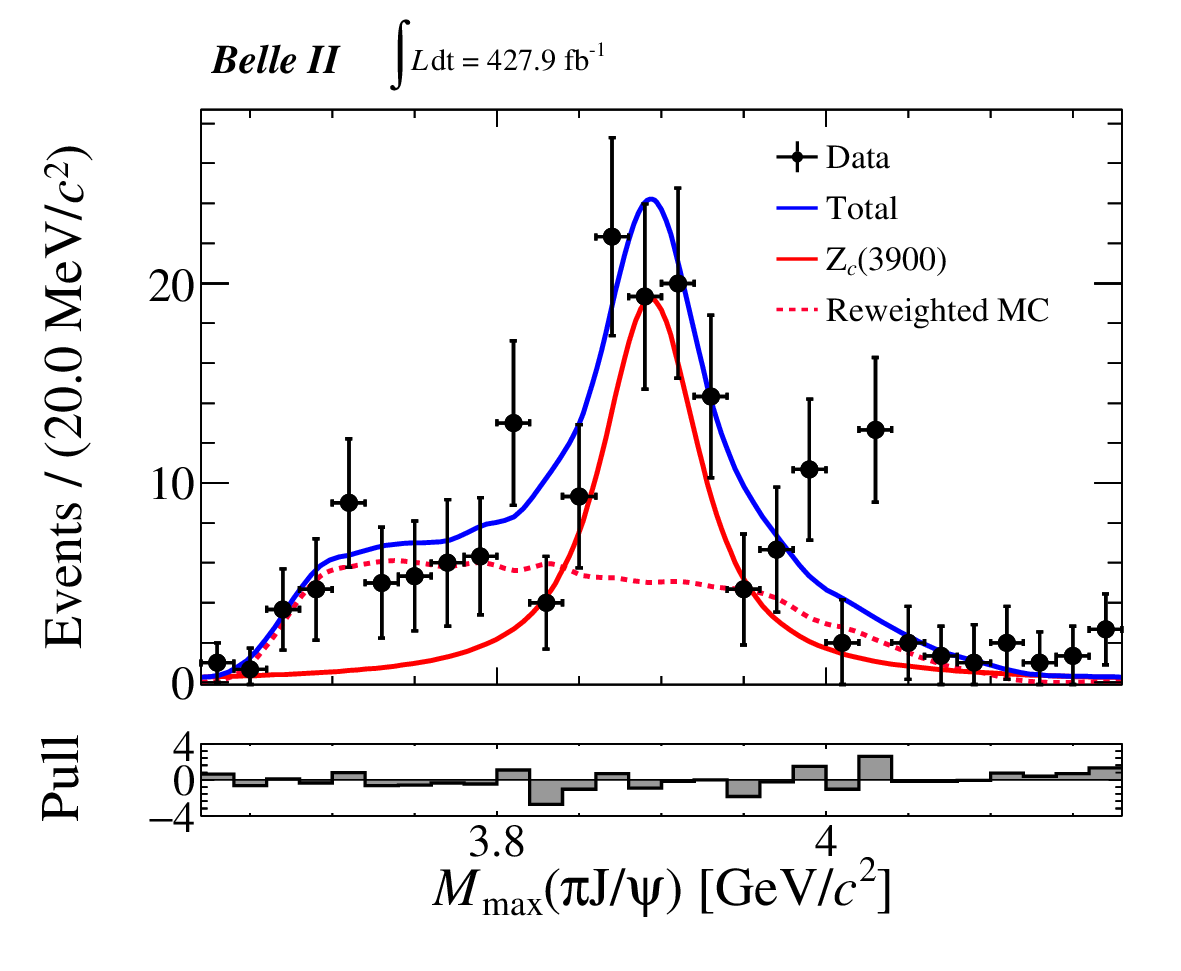}
    \put(-150, 120){\bf (a)} \\
    \includegraphics[width=7.0cm]{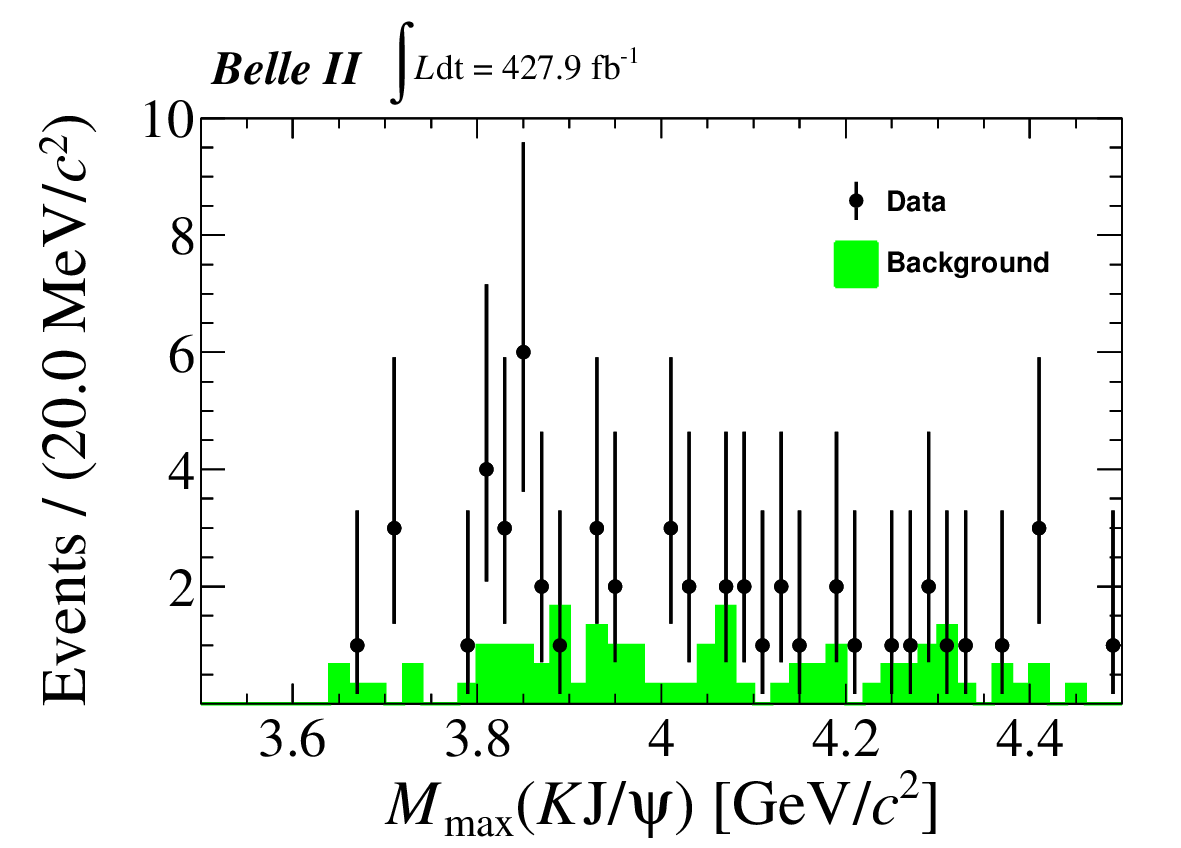}
    \put(-150, 100){\bf (b)} \\
    \caption{(a) An unbinned extended maximum likelihood fit to the distribution of the $M_{\rm max}(\pi\jpsi)$ in the $Y(4230/4320)$ region. (b) The distribution of the $M_{\rm max}(K\jpsi)$ with $M(\kkjpsi)<6.0~\gevcs$.}
    \label{fig:Zc}
\end{figure}

To investigate the band-like excesses in the $\pi^{\pm}\jpsi$ system, we fit the distribution of $M_{\max}(\pi\jpsi)$, the maximum of $M(\pi^+\jpsi)$ and $M(\pi^-\jpsi)$~\cite{Belle:2011,Liu:2013}. Figure~\ref{fig:Zc} shows the distribution of $M_{\rm max}(\pi\jpsi)$ in the $Y(4230/4320)$ region. An unbinned extended maximum likelihood fit is performed. The signal shape is parametrized as a Breit-Wigner function to describe $Z_c(3900)^{\pm}$, also known as $T_{c\bar{c}}(3900)^{\pm}$~\cite{PDG:2024}, whose parameters are fixed using the previous result from Belle~\cite{Liu:2013}. The signal shape is convolved with a Gaussian resolution function whose width is floating. The effect of cross-feed from the $\jpsi$ combined with a wrong $\pi$ on the signal shape has been checked and found to be negligible. The background from fake $\jpsi$ events is estimated using the normalized $\jpsi$ mass sideband regions. This background is fixed in the fit but is subtracted in Fig.~\ref{fig:Zc}. Other contributions, such as those from PHSP $\pipijpsi$ events and processes involving intermediate states like $f_0(980)\jpsi$, are estimated using a $\pipijpsi$ $M(\pipi)$-reweighted MC shape. This MC shape is based on the measured distribution from BESIII~\cite{BESIII:2025qkn} at $\sqrt{s}=4.26~{\rm GeV}$ and is rescaled according to the available phase space. The fitted resolution is consistent with the MC simulation. The statistical significance is calculated by comparing the logarithmic likelihoods with and without the $Z_c(3900)^{\pm}$ signal, and including the change of the number of parameters in the fits~\cite{Wilks:1938dza}. The significance of the $Z_{c}(3900)^{\pm}$ state is found to be $5.6\sigma$ in the nominal fit.

We account for systematic uncertainties from the background estimation and from the signal shape in the fit results by using different fit configurations. A set of alternative fits is performed to estimate the uncertainties, including using the $Z_c(3900)^{\pm}$ resonant parameters measured by BESIII~\cite{Ablikim:2013}, modifying the background shape with a second-order polynomial or redefining the sideband regions, and accounting for the uncertainty associated with a floating mass resolution. We also use the $\pipijpsi$ PHSP MC samples and perform an alternative fit, to estimate the uncertainty due to possible intermediate states in the $\pipi$ system such as the $f_0(980)$ state. The largest contribution to the systematic uncertainties arises from the resonance parameters. In all instances, the statistical significance of the $Z_c(3900)^{\pm}$ state remains above $5.3\sigma$. The Belle~II detector has better resolution and higher trigger efficiency than the Belle detector, leading to higher selection efficiency for ISR events~\cite{Belle-II:2018jsg}. Additionally, we use prior knowledge to fix the resonance parameters in the fit~\cite{Liu:2013}. As a result, our analysis reaches a significance for the $Z_c(3900)^{\pm}$ state comparable to Belle's, despite using a smaller data sample.

In $\EE\to\kkjpsi$, the distribution of $M_{\rm max}(K\jpsi)$ is presented in Fig.~\ref{fig:Zc} and no obvious structure is found in the $K^{\pm}\jpsi$ system. The limited size of the sample impedes further extraction of information regarding the three-body dynamics.

% -----------------------------------------
%     Cross section and uncertainties
% -----------------------------------------
\section{Cross section measurement}
The energy-dependent cross sections of $\EE\to\hhjpsi$ for $h=\pi/K/p$, from 3.8 GeV or the production threshold to 5.5/6.0/7.0 GeV in steps of 20/40/100 MeV, are calculated using
\begin{equation}
    \sigma_i = \frac{N_i^{\rm sig}}{\LK^{\rm eff}_i\cdot\eff_i\cdot\BF(\jpsill)}.
\end{equation}
Here, $N_i^{\rm sig} = \max(0,~n_i^{\rm obs}-f\times n_i^{\rm bkg})$~\cite{BESIII:2021,BelleII:2023}, where $n_i^{\rm obs}$, $n_i^{\rm bkg}$, and $f = \frac{1}{3}$ are the number of observed events in $\jpsi$ signal region in data, the number of background events determined from the $\jpsi$ sidebands extracted from counting, and the ratio between the numbers of background events in the $J/\psi$ signal region and sideband regions, respectively; $\LK^{\rm eff}_i$ is the effective ISR integrated luminosity obtained from the QED calculation~\cite{Kuraev:1985} in the $i$-th energy bin; $\BF(\jpsill)=11.93\%$ is taken from world average value~\cite{PDG:2024}. $\eff_i$ is the detection efficiency for the processes $\hhjpsi$ corrected for PID and trigger efficiencies. 

To estimate the detection efficiency, the binned efficiencies obtained from the $\pipijpsi$ PHSP signal MC samples are reweighted according to the Dalitz plot distributions measured from data in different energy regions. Specifically, the binned efficiency in each Dalitz plot bin $i$ is defined as $\varepsilon_i = N^{\mathrm{reco}}_i / N^{\mathrm{truth}}_i$, where $N^{\mathrm{reco}}_i$ and $N^{\mathrm{truth}}_i$ are the reconstructed and generated signal yields in bin $i$, respectively. Due to the limited sample at some energy points, different binning schemes are adopted to ensure stable efficiency estimates: a $5\times5$ binning is used at $\sqrt{s} = 4.26~\mathrm{GeV}$, while a $3\times3$ binning is used at other energies, in order to avoid negative yields after background subtraction. The data-driven detection efficiency is then calculated as $\varepsilon_{\mathrm{data}} = \sum_i \varepsilon_i N^{\mathrm{sig}}_i / N^{\mathrm{sig}}$, where $N^{\mathrm{sig}}_i$ is the number of signal events in data in bin $i$, and $N^{\mathrm{sig}}$ is the total number of signal events after background subtraction. The background contribution is estimated using the $J/\psi$ mass sideband. In particular, the $\pipijpsi$ MC sample at $\sqrt{s} = 3.686~{\rm GeV}$ uses the dynamic factor $f = (m_{\pipi}^2 - 4 m_{\pi}^2)^2$~\cite{BES:1999guu} for the $\pipi$ system to model the decay $\psip\to\pipijpsi$, by reweighting the $\pipi$ invariant mass distribution. The $\kkjpsi$ signal MC samples are reweighted using the $M(\kk)$ distribution from BESIII at $\sqrt{s} = 4.68~\mathrm{GeV}$~\cite{KKJpsi:2023} and rescaled according to the variation of size of kinematic phase-space across different energy points, by assuming that the dynamics do not change with $\sqrt{s}$. For the $\ppjpsi$ channel, the signal MC samples retain the PHSP model due to the lack of experimental measurements. The efficiency of each bin $\eff_i$ is obtained by fitting the efficiencies at eight energy points with a third-order polynomial function and then extrapolating the resulting efficiency curve for each channel. The systematic uncertainty arising from the $\hhjpsi$ dynamics is discussed in a later section.

While the comparison of the Dalitz plot for $\psip\to\pipijpsi$ between the signal MC sample and data shows good agreement, a significant discrepancy is observed in the $\cos\theta$ distribution of charged tracks in the control channel $\EE\to\gamma_{\rm ISR}\psip~(\to\pipijpsi)$ between data and MC samples in the end-cap regions, where large discrepancies in PID efficiency between data and MC simulation are known to exist. A data-driven method utilizing the control channel of $\EE\to\gamma_{\rm ISR}\psip~(\to\pipijpsi)$ is employed to reduce the data-MC discrepancy. The difference between data and MC samples is measured in different polar angle regions for electrons, muons, and pions, and the MC samples are reweighted to match the data. Since the events are well simulated in the barrel region, we scale the MC distributions by the ratio of data to MC yields in the central region ($-0.4 < \cos\theta < 0.6$ in the lab frame). Due to the limited size of the control sample, we first apply a one-dimensional (1D) reweighting to the $\cos\theta_{\ell}$~($\ell=e/\mu$) distributions. The overall reweighting factors for $\mu^+\mu^-$ and $e^+e^-$ modes are 0.874 and 0.819. After applying the correction factor for the $\cos\theta_{\ell}$ distributions, we then perform a 1D reweighting on the $\cos\theta_{\pi}$ distributions. The second reweighting factors for $\mu^+\mu^-$ and $e^+e^-$ modes are 0.967 and 0.961. A momentum-dependent correction is not considered due to limited statistics and weak $M(\pipijpsi)$ dependence of the track momentum distributions in the mass region of interest. After performing two 1D reweightings, we examined the $\cos\theta_\ell$, $\cos\theta_\pi$ and other kinematic distributions and found them to be consistent between data and MC simulations.

\begin{figure}[htpb]
    \centering
    \includegraphics[width=8.5cm]{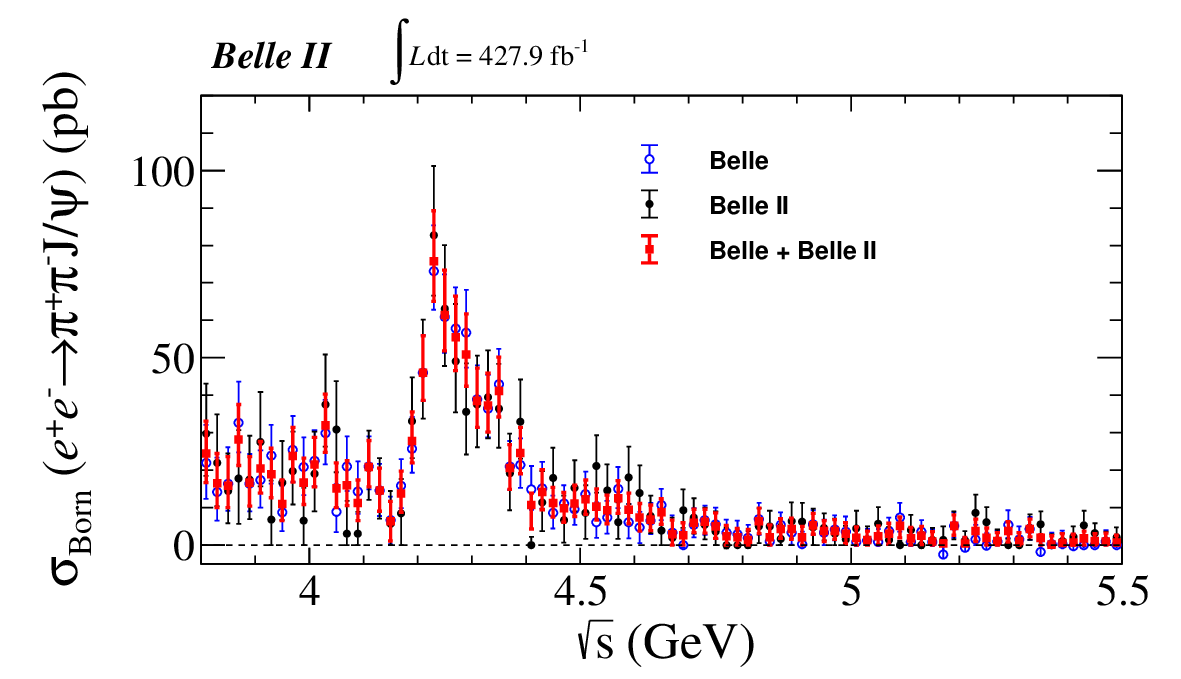}
    \put(-50, 101){\bf (a)} \\
    \includegraphics[width=8.5cm]{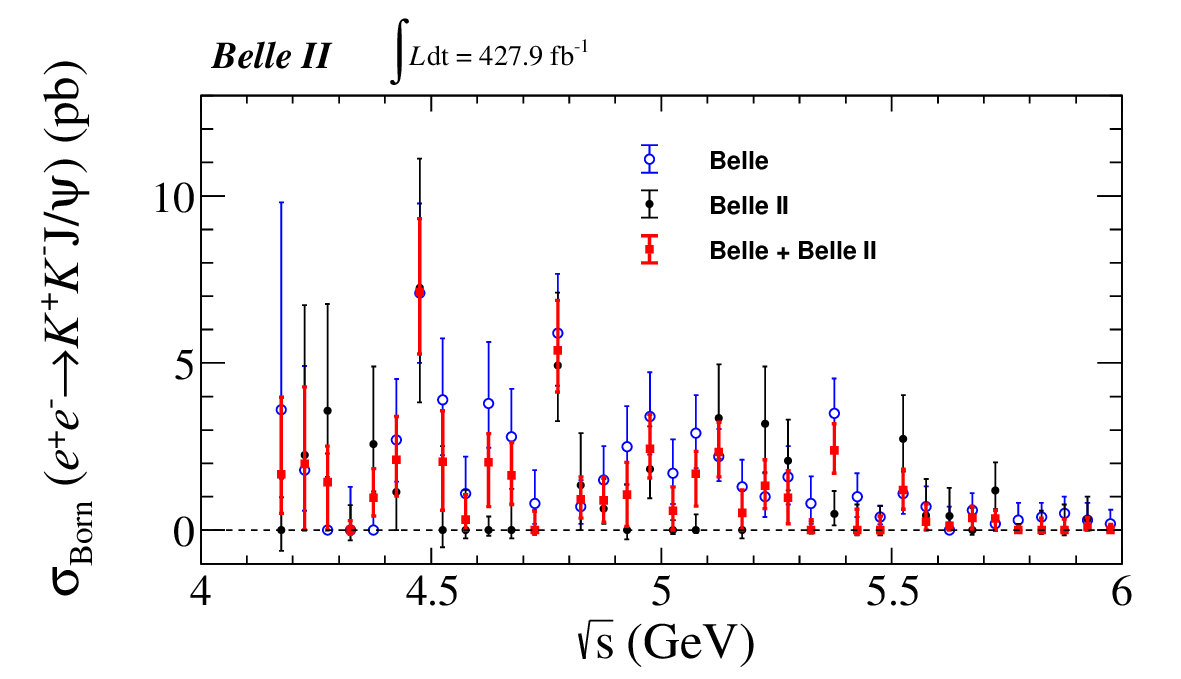}
    \put(-50, 101){\bf (b)} \\
    \includegraphics[width=8.5cm]{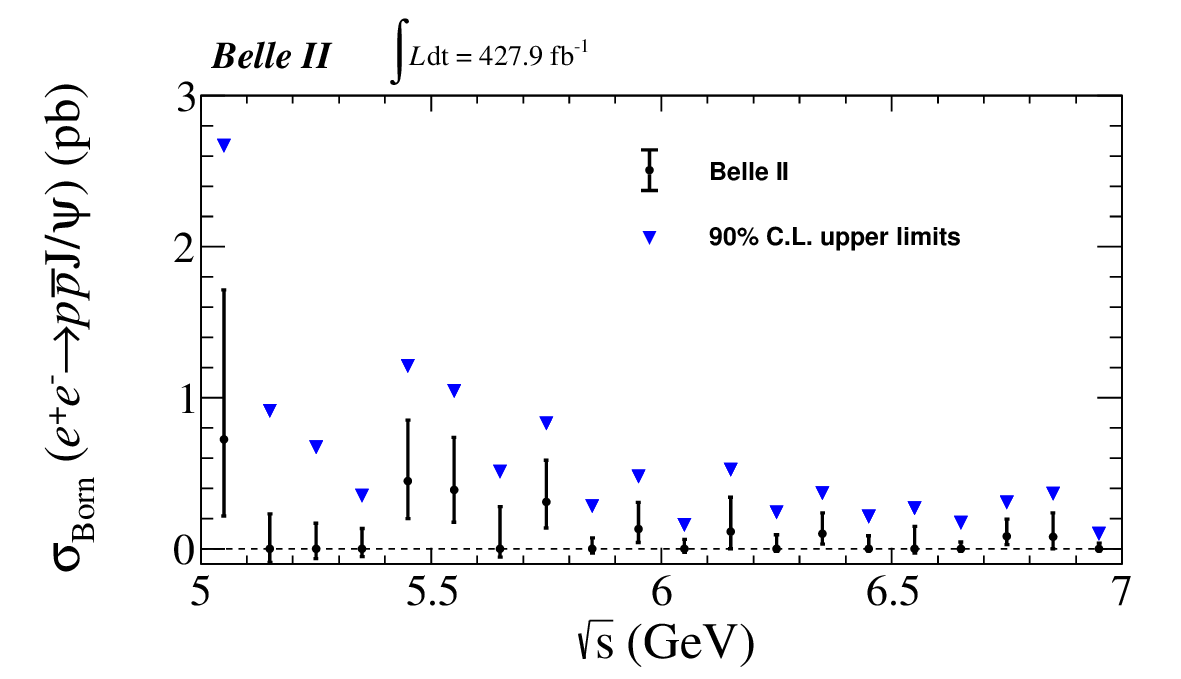}
    \put(-50, 101){\bf (c)}
    \caption{The measured Born cross sections of (a) $\EE\to\pipijpsi$ for $\ecms$ between 3.8 and 5.5 GeV, (b) $\EE\to\kkjpsi$ for $\ecms$ between 4.0 and 6.0 GeV, and (c) $\EE\to\ppjpsi$ for $\ecms$ between 5.0 and 7.0 GeV. The error bars include both statistical and systematic uncertainties. The blue open circles with error bars in panels (a) and (b) represent the Belle results~\cite{Liu:2013,Shen:2014}, while the red full squares with error bars indicate the combined results. The blue inverted triangles (c) are 90\% C.L. upper limits for the measured cross sections of $\EE\to\ppjpsi$.}
    \label{fig:xsec}
\end{figure}

The 68.3\% central confidence interval on $N_i^{\rm sig}$ for the $\hhjpsi$ measurements, and the additional 90\% C.L. upper limit on $N_i^{\rm sig}$ for the $\ppjpsi$ measurement, are estimated using the TRolke method~\cite{Brun:1997pa,Rolke:2005}, a frequentist approach that includes systematic uncertainties (Sect.~\ref{sec:syst}). Figure~\ref{fig:xsec} shows the calculated cross sections and the upper limits of cross sections for $\EE\to\hhjpsi$, where the error bars include both statistical and systematic uncertainties. Utilizing the combined number of signal and background events measured by Belle~\cite{Liu:2013,Shen:2014}, along with the total luminosity and the efficiency weighted by luminosity, we calculate combined cross sections and associated uncertainties at the 68.27\% C.L. for $\sigma_i$. Figure~\ref{fig:xsec} also presents the previous results from Belle~\cite{Liu:2013,Shen:2014} along with the combined results.

\section{Systematic uncertainties}\label{sec:syst}
The sources of systematic uncertainty associated with the cross section measurements are listed in Table~\ref{tab:syst}. The uncertainty in tracking efficiency is measured to be 0.3\% per track based on studies using control samples of $\EE\to\tau^+\tau^-$, where one tau lepton decays leptonically ($\tau^{\pm}\to\ell^{\pm}\nu_{\ell}\bar{\nu}_{\tau},~\ell=e,\mu$) while the other decays hadronically into three charged pions ($\tau^{\mp}\to3\pi^{\mp}\nu_{\tau}n\pi^{0},~n\geq0$). For the four-track final states in this analysis, the total uncertainty is therefore 1.2\%. The trigger efficiencies in the CDC, ECL and KLM are evaluated with the control samples $\EE\to\gamma_{\rm ISR}\psip$ and $\EE\to\gamma\MM$, with the statistical uncertainty of 0.7\% for the correction factor included as the trigger uncertainty. The integrated luminosity contributes an uncertainty of 0.5\%~\cite{luminosity:2025}. The uncertainty for $\BF(\jpsill)$ is considered to be 0.4\%, calculated by combining the errors of the world averages for the $\EE$ and $\MM$ modes~\cite{PDG:2024}.

The statistical uncertainty of the reweighting correction factor is calculated to be 6.5\%, and considered as the $\cos\theta$-related uncertainty. The uncertainty associated with this method can be reduced by increasing the size of the control sample in the future. The systematic uncertainties of PID efficiency in the central region are estimated from data and MC samples of leptons and hadrons in $\jpsi\to\LL$, $\EE\to\EE\LL$, $\EE\to \tau^\pm(\to3\pi^\pm\nu_\tau)\tau^\mp(\to\ell^\mp\bar{\nu}_{\ell}\nu_{\tau})$, $\EE\to\MM\gamma$, $K^0_S\to\pipi$, $D^{*+}\to \pi^+ D^0~(\to K^-\pi^+)$, and $\Lambda\to p\pi^-$~\cite{syscorrfw}. The uncertainties in the total electron, muon, pion, kaon, and proton identification efficiencies are estimated to be approximately 0.7\%, 0.2\%, 0.1\%, 3.0\%, and 1.3\%, respectively, taking into account the momentum and polar angle dependence of the tracks~\cite{Belle-II:2025tpe}. The overall PID uncertainties are reported as 0.7\%, 3.1\%, and 1.5\% for $\pipijpsi$, $\kkjpsi$ and $\ppjpsi$, respectively.

The uncertainties arising from the MC generator are related to the dynamics of three-body decays. The uncertainty arising from the modeling of the $M(\pipi)$ and $M(\kk)$ distributions is estimated by comparing with the corresponding PHSP spectrum models. The uncertainty related to the modeling of the $M(\pp)$ distribution is evaluated by comparing with a reweighted $M(\pp)$ spectrum model, which is constructed based on the $M(\pipi/\kk)$ results from BESIII and rescaled according to the available size of kinematic phase-space~\cite{BESIII:2025qkn,KKJpsi:2023}. The resulting efficiencies are higher by 1\% to 7\%, depending on the energy point. We use the maximum deviation across all energies in each channel to quote conservative uncertainties of 4\%, 4\%, and 7\% for $\pipijpsi$, $\kkjpsi$, and $\ppjpsi$, respectively. The MC statistical error on the efficiency is 0.3\%. To evaluate the uncertainty arising from the background estimation, we varied the sideband regions used for background determination. The largest difference in the background level among these variations is taken as the uncertainty, which is 3.6\%, 4.3\%, and 9.5\% for the channels $\pipijpsi$, $\kkjpsi$ and $\ppjpsi$, respectively.

Assuming that all sources except for the background estimation are independent, they are combined in quadrature to yield the total multiplicative systematic uncertainty associated with the cross section measurements. The uncertainty arising from background estimation is treated as additive. It is incorporated by varying the number of background events by $\pm1\sigma$ and taking the largest resulting uncertainty as a conservative estimate within the TRolke statistical method. The final uncertainty thus includes statistical, additive and multiplicative systematic uncertainties. For the combined results, we combine the multiplicative systematic uncertainties from Belle and Belle~II in quadrature, assuming the systematic uncertainties are uncorrelated. This assumption has a negligible impact because the measurements from the two experiments are mainly limited by statistical uncertainties.

\begin{table}[htpb]
    \centering
    \caption{The multiplicative relative systematic uncertainties in the cross section measurements of $\EE\to\hhjpsi$. Uncertainties are quoted in percent and are common for all data points.}
    \begin{tabular}{lccc}
        \hline\hline
        Source & $\pipijpsi$ & $\kkjpsi$ & $\ppjpsi$ \\
        \hline
        Tracking & 1.2 & 1.2 & 1.2 \\
        Trigger & 0.7 & 0.7 & 0.7 \\
        Luminosity & 0.5 & 0.5 & 0.5 \\
        $\BF(\jpsill)$ & 0.4 & 0.4 & 0.4 \\
        $\cos\theta$ & 6.5 & 6.5 & 6.5 \\
        PID & 0.7 & 3.1 & 1.5 \\
        $M(\hh)$ modeling & 4.0 & 4.0 & 7.0 \\
        MC sample size & 0.3 & 0.3 & 0.3 \\
        \hline
        Total multiplicative & 7.8 & 8.4 & 9.8 \\
        \hline\hline
    \end{tabular}
    \label{tab:syst}
\end{table}

% -----------------------
%     Control samples
% -----------------------
To validate our analysis, we also measured the ISR $\psip$ production rate using the same selection criteria. The cross sections are determined to be $(13.8\pm0.2\pm0.9)~{\rm pb}$ and $(14.8\pm0.2\pm1.0)~{\rm pb}$ at $\ecms=10.58~\gev$ for the $\MM$ and $\EE$ modes, respectively. The systematic uncertainties here, except for those from the $M(\hh)$ modeling, are the same as those listed in Table~\ref{tab:syst}. The uncertainty in the $M(\hh)$ modeling is estimated by varying the dynamic factor for the $\pipi$ system. An additional uncertainty from $\BF[\psip\to\pipijpsi]$ is taken from the world averages~\cite{PDG:2024}. These measurements are consistent within uncertainties with the predicted value of $(13.6\pm0.4)~{\rm pb}$ at $\ecms=10.58~\gev$ from Ref.~\cite{Kuraev:1985}, as well as with previous measurements~\cite{Yuan:2007,Liu:2013}.

% --------------------
%     Combination
% --------------------
Tables~\ref{tab:pi_xsec}, \ref{tab:k_xsec}, and \ref{tab:p_xsec} summarize the final results along with all the relevant information used in the cross section measurements or upper limit calculations for $\EE \to \pipijpsi$, $\EE \to \kkjpsi$, and $\EE \to \ppjpsi$, respectively. The combined results from Belle~\cite{Liu:2013,Shen:2014} and Belle~II measurements are presented in Tables~\ref{tab:pi_comb_xsec} and \ref{tab:k_comb_xsec}.

% ------------------------------
%     Summary and Discussion
% ------------------------------
\section{Summary}
In summary, the cross sections for the processes $\EE \to \hhjpsi$ at c.m.\ energies ranging from 3.8 GeV or the production threshold up to 5.5/6.0/7.0 GeV for $h=\pi/K/p$ are measured using a $427.9~\mathrm{fb}^{-1}$ data sample collected by Belle~II. The total cross sections and their uncertainties are calculated to compare the results from different experiments. The measured cross sections of $\EE \to \pipijpsi$ and $\EE \to \kkjpsi$ agree with previous experimental results within $1\sigma$ and $2\sigma$ uncertainties, respectively. The combined cross sections from Belle and Belle~II differ from the BESIII results~\cite{BESIII:2017,BESIII:2022} by less than $1\sigma$. The measurement precision is comparable to that achieved by BaBar~\cite{BaBar:2012} and Belle using ISR. However, the measurement is less precise than the BESIII results obtained from dedicated ``XYZ'' energy scans, which have a total integrated luminosity of about $500~{\rm pb}^{-1}$ at each energy point with energy steps of approximately 20 MeV, yielding precisions of about 5\% and 10\% for the $\pipijpsi$ and $\kkjpsi$ processes, respectively. The corresponding effective ISR luminosity at Belle~II over the same energy range and step size is only 30 to 50 ${\rm pb}^{-1}$.

In the $\pipijpsi$ invariant mass distribution, an enhancement around 4.26 GeV is seen, indicative of the $Y(4230/4320)$ state. A global coupled-channel analysis~\cite{Nakamura:2023obk} suggests a possible $\psi(4040)$-like enhancement in the $M(\pipijpsi)$ distribution, and we find a small excess near 4.1 GeV with a statistical significance of $2.0\sigma$. For the $\kkjpsi$ invariant mass distribution, no significant structure is found, although the statistical power of the data is poor. For the first time, the process $\EE \to \ppjpsi$ is investigated, but the yield is not significant and no clear structure is observed in the $\ppjpsi$ cross section. Upper limits on the cross sections for $\EE \to \ppjpsi$ from the production threshold up to 7.0 GeV are established. Our result near 6 GeV ($<0.16~\mathrm{pb}$) is consistent with the theoretical prediction at ${\cal O}(4~\mathrm{fb})$~\cite{Zhang:2025pfz}, and is expected to provide valuable input for future theoretical studies. A clear signal with a significance of $5.3\sigma$ for the $Z_c(3900)^{\pm}$ is observed in the $\pi^{\pm}\jpsi$ system; however, no structure is found in the $K^{\pm}\jpsi$ system. These results are consistent with previous measurements performed by Belle and BESIII~\cite{Yuan:2007,Liu:2013,KKJpsi:2022,KKJpsi:2023}. A larger Belle~II data sample will be crucial for further investigations of the structures in the $h^{\pm}\jpsi$ and $\hhjpsi$ systems, which may help clarify the nature of these exotic states.

%%%%%%%%%%%%%%% Appendix %%%%%%%%%%%%%%%
\begin{table*}[htbp]
    \centering
    \caption{Summary of cross sections ($\sigma$) of $\EE\to\pipijpsi$. We also list the $\EE$ c.m.\ energy ($\ecms$), number of observed events ($n^{\rm obs}$), number of backgrounds estimated from the normalized $\jpsi$ mass sidebands ($n^{\rm bkg}$), detection efficiency ($\varepsilon$), and effective ISR luminosity ($\lum$). All values are calculated for a 20 MeV bin size and $\ecms$ is the central value of the bin. The uncertainties include both statistical and systematic contributions.}
    \begin{tabular}{lcccccccccccc}
        \hline\hline
        $\ecms$ (GeV) & $n^{\rm obs}$ & $n^{\rm bkg}$ & $\varepsilon~(\%)$ & $\lum~({\rm pb}^{-1})$ & $\sigma~(\rm pb)$ && $\ecms$ (GeV) & $n^{\rm obs}$ & $n^{\rm bkg}$ & $\varepsilon~(\%)$ & $\lum~({\rm pb}^{-1})$ & $\sigma~(\rm pb)$ \\
        \hline
3.810 & 9 & 1 & 7.1 & 31.9 & $29.8^{+13.4}_{-11.3}$ && 4.670 & 4 & 3 & 10.4 & 42.8 & $1.9^{+5.4}_{-1.9}$\\
3.830 & 8 & 2 & 7.1 & 32.2 & $22.0^{+12.9}_{-11.5}$ && 4.690 & 6 & 1 & 10.5 & 43.1 & $9.3^{+5.6}_{-4.7}$\\
3.850 & 5 & 1 & 7.2 & 32.4 & $14.4^{+10.1}_{-8.6}$ && 4.710 & 5 & 1 & 10.6 & 43.4 & $7.3^{+5.1}_{-4.4}$\\
3.870 & 8 & 3 & 7.2 & 32.6 & $17.7^{+13.0}_{-12.1}$ && 4.730 & 4 & 1 & 10.7 & 43.7 & $5.4^{+4.6}_{-4.0}$\\
3.890 & 7 & 2 & 7.3 & 32.8 & $17.5^{+11.7}_{-10.5}$ && 4.750 & 5 & 3 & 10.8 & 44.0 & $3.5^{+5.5}_{-3.5}$\\
3.910 & 10 & 2 & 7.4 & 33.1 & $27.5^{+13.4}_{-11.8}$ && 4.770 & 3 & 3 & 10.9 & 44.3 & $0.0^{+4.6}_{-0.9}$\\
3.930 & 5 & 3 & 7.4 & 33.3 & $6.8^{+10.5}_{-6.8}$ && 4.790 & 2 & 3 & 10.9 & 44.6 & $0.0^{+2.3}_{-0.5}$\\
3.950 & 7 & 2 & 7.5 & 33.5 & $16.7^{+11.2}_{-10.0}$ && 4.810 & 2 & 2 & 11.0 & 44.9 & $0.0^{+3.7}_{-0.7}$\\
3.970 & 7 & 1 & 7.6 & 33.8 & $19.7^{+10.5}_{-8.9}$ && 4.830 & 5 & 2 & 11.1 & 45.2 & $5.0^{+4.9}_{-4.5}$\\
3.990 & 4 & 2 & 7.6 & 34.0 & $6.5^{+8.8}_{-6.5}$ && 4.850 & 4 & 1 & 11.2 & 45.5 & $4.9^{+4.2}_{-3.6}$\\
4.010 & 8 & 2 & 7.7 & 34.2 & $19.1^{+11.2}_{-10.0}$ && 4.870 & 3 & 2 & 11.3 & 45.8 & $1.6^{+4.0}_{-1.6}$\\
4.030 & 13 & 1 & 7.8 & 34.5 & $37.5^{+13.3}_{-11.3}$ && 4.890 & 6 & 2 & 11.4 & 46.1 & $6.4^{+5.0}_{-4.5}$\\
4.050 & 12 & 2 & 7.9 & 34.7 & $30.8^{+12.9}_{-11.4}$ && 4.910 & 6 & 2 & 11.5 & 46.4 & $6.3^{+4.9}_{-4.5}$\\
4.070 & 4 & 3 & 7.9 & 34.9 & $3.0^{+8.7}_{-3.0}$ && 4.930 & 6 & 3 & 11.6 & 46.7 & $4.6^{+5.1}_{-4.6}$\\
4.090 & 2 & 1 & 8.0 & 35.2 & $3.0^{+5.9}_{-3.0}$ && 4.950 & 4 & 1 & 11.7 & 47.1 & $4.6^{+3.9}_{-3.3}$\\
4.110 & 8 & 1 & 8.1 & 35.4 & $20.5^{+10.0}_{-8.4}$ && 4.970 & 4 & 2 & 11.8 & 47.4 & $3.0^{+4.1}_{-3.0}$\\
4.130 & 6 & 1 & 8.1 & 35.6 & $14.4^{+8.7}_{-7.4}$ && 4.990 & 3 & 2 & 11.9 & 47.7 & $1.5^{+3.6}_{-1.5}$\\
4.150 & 5 & 3 & 8.2 & 35.9 & $5.7^{+8.8}_{-5.7}$ && 5.010 & 6 & 3 & 12.1 & 48.0 & $4.3^{+4.8}_{-4.3}$\\
4.170 & 6 & 3 & 8.3 & 36.1 & $8.4^{+9.2}_{-8.4}$ && 5.030 & 3 & 2 & 12.2 & 48.4 & $1.4^{+3.5}_{-1.4}$\\
4.190 & 13 & 1 & 8.4 & 36.4 & $33.1^{+11.7}_{-9.9}$ && 5.050 & 6 & 2 & 12.3 & 48.7 & $5.6^{+4.4}_{-4.0}$\\
4.210 & 19 & 2 & 8.4 & 36.6 & $46.1^{+14.1}_{-12.3}$ && 5.070 & 3 & 2 & 12.4 & 49.0 & $1.4^{+3.4}_{-1.4}$\\
4.230 & 33 & 2 & 8.5 & 36.9 & $82.7^{+18.5}_{-16.2}$ && 5.090 & 1 & 2 & 12.5 & 49.4 & $0.0^{+1.1}_{-0.2}$\\
4.250 & 28 & 4 & 8.6 & 37.1 & $63.0^{+17.0}_{-15.2}$ && 5.110 & 5 & 2 & 12.6 & 49.7 & $4.0^{+3.9}_{-3.6}$\\
4.270 & 23 & 4 & 8.7 & 37.4 & $49.1^{+15.2}_{-13.7}$ && 5.130 & 3 & 3 & 12.7 & 50.1 & $0.0^{+3.5}_{-0.7}$\\
4.290 & 17 & 3 & 8.8 & 37.6 & $35.6^{+12.8}_{-11.4}$ && 5.150 & 3 & 2 & 12.8 & 50.4 & $1.3^{+3.2}_{-1.3}$\\
4.310 & 18 & 3 & 8.8 & 37.9 & $37.6^{+13.0}_{-11.5}$ && 5.170 & 4 & 3 & 12.9 & 50.8 & $1.3^{+3.7}_{-1.3}$\\
4.330 & 18 & 2 & 8.9 & 38.1 & $39.4^{+12.5}_{-10.9}$ && 5.190 & 6 & 2 & 13.0 & 51.1 & $5.0^{+4.0}_{-3.6}$\\
4.350 & 17 & 2 & 9.0 & 38.4 & $36.4^{+12.0}_{-10.5}$ && 5.210 & 2 & 1 & 13.1 & 51.5 & $1.2^{+2.5}_{-1.2}$\\
4.370 & 12 & 4 & 9.1 & 38.7 & $19.1^{+10.6}_{-9.8}$ && 5.230 & 10 & 3 & 13.2 & 51.8 & $8.6^{+4.9}_{-4.5}$\\
4.390 & 16 & 2 & 9.2 & 38.9 & $32.9^{+11.3}_{-9.9}$ && 5.250 & 7 & 2 & 13.4 & 52.2 & $6.0^{+4.0}_{-3.6}$\\
4.410 & 2 & 5 & 9.2 & 39.2 & $0.0^{+2.2}_{-0.4}$ && 5.270 & 3 & 2 & 13.5 & 52.5 & $1.2^{+2.9}_{-1.2}$\\
4.430 & 8 & 3 & 9.3 & 39.5 & $11.4^{+8.3}_{-7.7}$ && 5.290 & 2 & 3 & 13.6 & 52.9 & $0.0^{+1.6}_{-0.3}$\\
4.450 & 9 & 1 & 9.4 & 39.7 & $17.9^{+8.1}_{-6.8}$ && 5.310 & 3 & 4 & 13.7 & 53.3 & $0.0^{+2.1}_{-0.4}$\\
4.470 & 5 & 2 & 9.5 & 40.0 & $6.6^{+6.5}_{-6.0}$ && 5.330 & 6 & 2 & 13.8 & 53.7 & $4.5^{+3.6}_{-3.2}$\\
4.490 & 8 & 1 & 9.6 & 40.3 & $15.2^{+7.4}_{-6.2}$ && 5.350 & 6 & 1 & 13.9 & 54.0 & $5.6^{+3.4}_{-2.8}$\\
4.510 & 7 & 3 & 9.7 & 40.5 & $8.6^{+7.5}_{-7.0}$ && 5.370 & 3 & 3 & 14.0 & 54.4 & $0.0^{+2.9}_{-0.6}$\\
4.530 & 11 & 1 & 9.8 & 40.8 & $21.1^{+8.3}_{-7.0}$ && 5.390 & 2 & 1 & 14.2 & 54.8 & $1.1^{+2.1}_{-1.1}$\\
4.550 & 8 & 1 & 9.8 & 41.1 & $14.5^{+7.1}_{-6.0}$ && 5.410 & 4 & 2 & 14.3 & 55.2 & $2.1^{+2.9}_{-2.1}$\\
4.570 & 4 & 1 & 9.9 & 41.4 & $6.1^{+5.2}_{-4.5}$ && 5.430 & 8 & 3 & 14.4 & 55.6 & $5.2^{+3.8}_{-3.6}$\\
4.590 & 11 & 2 & 10.0 & 41.6 & $18.1^{+8.1}_{-7.1}$ && 5.450 & 4 & 1 & 14.5 & 56.0 & $3.1^{+2.6}_{-2.3}$\\
4.610 & 9 & 2 & 10.1 & 41.9 & $13.8^{+7.3}_{-6.5}$ && 5.470 & 3 & 2 & 14.6 & 56.4 & $1.0^{+2.5}_{-1.0}$\\
4.630 & 5 & 1 & 10.2 & 42.2 & $7.8^{+5.4}_{-4.6}$ && 5.490 & 4 & 2 & 14.8 & 56.8 & $2.0^{+2.7}_{-2.0}$\\
4.650 & 4 & 2 & 10.3 & 42.5 & $3.8^{+5.2}_{-3.8}$ && & & & & & \\
        \hline\hline
    \end{tabular}
    \label{tab:pi_xsec}
\end{table*}

\begin{table*}[htbp]
    \centering
    \caption{Summary of cross sections ($\sigma$) of $\EE\to\kkjpsi$. We also list the $\EE$ c.m.\ energy ($\ecms$), number of observed events ($n^{\rm obs}$), number of backgrounds estimated from the normalized $\jpsi$ mass sidebands ($n^{\rm bkg}$), detection efficiency ($\varepsilon$), and effective ISR luminosity ($\lum$). All values are calculated for a 50 MeV bin size and $\ecms$ is the central value of the bin. The uncertainties include both statistical and systematic contributions.}
    \begin{tabular}{lcccccccccccc}
        \hline\hline
        $\ecms$ (GeV) & $n^{\rm obs}$ & $n^{\rm bkg}$ & $\varepsilon~(\%)$ & $\lum~({\rm pb}^{-1})$ & $\sigma~(\rm pb)$ && $\ecms$ (GeV) & $n^{\rm obs}$ & $n^{\rm bkg}$ & $\varepsilon~(\%)$ & $\lum~({\rm pb}^{-1})$ & $\sigma~(\rm pb)$ \\
        \hline
4.175 & 0 & 0 & 3.0 & 90.5 & $0.00^{+1.56}_{-0.62}$ && 5.125 & 6 & 0 & 12.0 & 124.9 & $3.35^{+1.61}_{-1.20}$\\
4.225 & 2 & 1 & 4.0 & 92.0 & $2.26^{+4.48}_{-2.26}$ && 5.175 & 2 & 2 & 12.1 & 127.1 & $0.00^{+1.19}_{-0.24}$\\
4.275 & 2 & 0 & 5.0 & 93.6 & $3.58^{+3.20}_{-1.98}$ && 5.225 & 7 & 1 & 12.2 & 129.3 & $3.19^{+1.72}_{-1.45}$\\
4.325 & 0 & 1 & 5.9 & 95.2 & $0.00^{+0.75}_{-0.30}$ && 5.275 & 4 & 0 & 12.3 & 131.6 & $2.07^{+1.24}_{-0.88}$\\
4.375 & 2 & 0 & 6.7 & 96.8 & $2.58^{+2.31}_{-1.43}$ && 5.325 & 0 & 1 & 12.4 & 133.9 & $0.00^{+0.25}_{-0.10}$\\
4.425 & 2 & 1 & 7.4 & 98.5 & $1.15^{+2.28}_{-1.15}$ && 5.375 & 1 & 0 & 12.5 & 136.3 & $0.49^{+0.67}_{-0.34}$\\
4.475 & 9 & 2 & 8.1 & 100.2 & $7.25^{+3.87}_{-3.43}$ && 5.425 & 1 & 1 & 12.6 & 138.7 & $0.00^{+0.77}_{-0.15}$\\
4.525 & 3 & 3 & 8.7 & 101.9 & $0.00^{+2.51}_{-0.50}$ && 5.475 & 1 & 1 & 12.7 & 141.1 & $0.00^{+0.75}_{-0.15}$\\
4.575 & 2 & 3 & 9.2 & 103.6 & $0.00^{+1.21}_{-0.24}$ && 5.525 & 6 & 0 & 12.8 & 143.7 & $2.73^{+1.31}_{-0.98}$\\
4.625 & 0 & 1 & 9.7 & 105.4 & $0.00^{+0.41}_{-0.16}$ && 5.575 & 3 & 2 & 13.0 & 146.2 & $0.44^{+1.09}_{-0.44}$\\
4.675 & 1 & 1 & 10.1 & 107.2 & $0.00^{+1.24}_{-0.25}$ && 5.625 & 2 & 1 & 13.2 & 148.9 & $0.43^{+0.85}_{-0.43}$\\
4.725 & 1 & 2 & 10.4 & 109.0 & $0.00^{+0.60}_{-0.12}$ && 5.675 & 1 & 1 & 13.4 & 151.6 & $0.00^{+0.66}_{-0.13}$\\
4.775 & 7 & 0 & 10.7 & 110.9 & $4.93^{+2.18}_{-1.66}$ && 5.725 & 3 & 0 & 13.7 & 154.3 & $1.19^{+0.84}_{-0.57}$\\
4.825 & 3 & 1 & 11.0 & 112.7 & $1.35^{+1.55}_{-1.35}$ && 5.775 & 0 & 1 & 14.0 & 157.2 & $0.00^{+0.19}_{-0.08}$\\
4.875 & 1 & 0 & 11.3 & 114.7 & $0.65^{+0.89}_{-0.45}$ && 5.825 & 1 & 1 & 14.3 & 160.1 & $0.00^{+0.58}_{-0.12}$\\
4.925 & 2 & 2 & 11.5 & 116.7 & $0.00^{+1.37}_{-0.27}$ && 5.875 & 2 & 2 & 14.7 & 163.0 & $0.00^{+0.76}_{-0.15}$\\
4.975 & 3 & 0 & 11.6 & 118.7 & $1.82^{+1.29}_{-0.87}$ && 5.925 & 2 & 1 & 15.1 & 166.1 & $0.33^{+0.66}_{-0.33}$\\
5.025 & 0 & 0 & 11.8 & 120.7 & $0.00^{+0.29}_{-0.12}$ && 5.975 & 0 & 0 & 15.6 & 169.2 & $0.00^{+0.16}_{-0.06}$\\
5.075 & 1 & 2 & 11.9 & 122.8 & $0.00^{+0.47}_{-0.09}$ && & & & & & \\
        \hline\hline
    \end{tabular}
    \label{tab:k_xsec}
\end{table*}

\begin{table*}[htbp]
    \centering
    \caption{Summary of cross sections ($\sigma$) and upper limit at 90\% C.L. of cross sections $\sigma^{\rm U.L.}$ of $\EE\to\ppjpsi$. We also list the $\EE$ c.m.\ energy ($\ecms$), number of observed events ($n^{\rm obs}$), number of backgrounds estimated from the normalized $\jpsi$ mass sidebands ($n^{\rm bkg}$), detection efficiency ($\varepsilon$), and effective ISR luminosity ($\lum$). All values are calculated for a 100 MeV bin size and $\ecms$ is the central value of the bin. The uncertainties include both statistical and systematic contributions.}
    \begin{tabular}{lcccccc}
        \hline\hline
        $\ecms$ (GeV) & $n^{\rm obs}$ & $n^{\rm bkg}$ & $\varepsilon~(\%)$ & $\lum~({\rm pb}^{-1})$ & $\sigma~(\rm pb)$ & $\sigma^{\rm U.L.}~(\rm pb)$ \\
        \hline
5.050 & 1 & 0 & 4.8 & 243.5 & $0.72^{+0.99}_{-0.50}$ & 2.67 \\
5.150 & 0 & 0 & 7.3 & 252.0 & $0.00^{+0.23}_{-0.09}$ & 0.91 \\
5.250 & 0 & 0 & 9.5 & 260.9 & $0.00^{+0.17}_{-0.07}$ & 0.67 \\
5.350 & 0 & 1 & 11.6 & 270.1 & $0.00^{+0.13}_{-0.05}$ & 0.35 \\
5.450 & 2 & 0 & 13.3 & 279.8 & $0.45^{+0.40}_{-0.25}$ & 1.21 \\
5.550 & 2 & 0 & 14.9 & 289.9 & $0.39^{+0.35}_{-0.21}$ & 1.04 \\
5.650 & 1 & 1 & 16.3 & 300.4 & $0.00^{+0.28}_{-0.06}$ & 0.51 \\
5.750 & 2 & 0 & 17.5 & 311.5 & $0.31^{+0.28}_{-0.17}$ & 0.83 \\
5.850 & 0 & 0 & 18.5 & 323.1 & $0.00^{+0.07}_{-0.03}$ & 0.28 \\
5.950 & 1 & 0 & 19.3 & 335.3 & $0.13^{+0.18}_{-0.09}$ & 0.48 \\
6.050 & 0 & 1 & 20.1 & 348.0 & $0.00^{+0.06}_{-0.02}$ & 0.16 \\
6.150 & 2 & 1 & 20.6 & 361.5 & $0.11^{+0.23}_{-0.11}$ & 0.52 \\
6.250 & 1 & 2 & 21.1 & 375.6 & $0.00^{+0.09}_{-0.02}$ & 0.24 \\
6.350 & 1 & 0 & 21.5 & 390.6 & $0.10^{+0.14}_{-0.07}$ & 0.37 \\
6.450 & 1 & 2 & 21.7 & 406.3 & $0.00^{+0.08}_{-0.02}$ & 0.21 \\
6.550 & 1 & 1 & 21.9 & 423.0 & $0.00^{+0.15}_{-0.03}$ & 0.27 \\
6.650 & 0 & 0 & 22.1 & 440.6 & $0.00^{+0.04}_{-0.02}$ & 0.17 \\
6.750 & 1 & 0 & 22.2 & 459.3 & $0.08^{+0.11}_{-0.06}$ & 0.30 \\
6.850 & 2 & 1 & 22.2 & 479.1 & $0.08^{+0.16}_{-0.08}$ & 0.37 \\
6.950 & 0 & 1 & 22.3 & 500.1 & $0.00^{+0.04}_{-0.02}$ & 0.10 \\
        \hline\hline
    \end{tabular}
    \label{tab:p_xsec}
\end{table*}

\begin{table*}[htbp]
    \centering
    \caption{Combined cross sections $\sigma$ of $\EE\to\pipijpsi$ of Belle~\cite{Liu:2013} and Belle~II measurements. We also list the $\EE$ c.m.\ energy ($\ecms$), number of observed events ($n^{\rm obs}$), number of backgrounds estimated from the normalized $\jpsi$ mass sidebands ($n^{\rm bkg}$), detection efficiency ($\varepsilon$), and effective ISR luminosity ($\lum$). All values are calculated for a 20 MeV bin size and $\ecms$ is the central value of the bin. The uncertainties include both statistical and systematic contributions.}
    \begin{tabular}{lcccccccccccc}
        \hline\hline
        $\ecms$ (GeV) & $n^{\rm obs}$ & $n^{\rm bkg}$ & $\varepsilon~(\%)$ & $\lum~({\rm pb}^{-1})$ & $\sigma~(\rm pb)$ && $\ecms$ (GeV) & $n^{\rm obs}$ & $n^{\rm bkg}$ & $\varepsilon~(\%)$ & $\lum~({\rm pb}^{-1})$ & $\sigma~(\rm pb)$ \\
        \hline
3.810 & 30 & 9 & 6.9 & 104.1 & $24.40^{+8.68}_{-7.80}$ && 4.670 & 19 & 13 & 11.1 & 139.5 & $2.87^{+3.76}_{-2.87}$\\
3.830 & 24 & 9 & 7.1 & 104.8 & $16.56^{+7.91}_{-6.55}$ && 4.690 & 18 & 13 & 11.2 & 140.4 & $2.66^{+3.29}_{-2.66}$\\
3.850 & 23 & 8 & 7.2 & 105.6 & $15.75^{+7.85}_{-5.76}$ && 4.710 & 21 & 9 & 11.3 & 141.4 & $5.96^{+3.65}_{-2.72}$\\
3.870 & 35 & 8 & 7.4 & 106.3 & $28.18^{+9.36}_{-6.91}$ && 4.730 & 21 & 9 & 11.3 & 142.4 & $6.23^{+3.26}_{-3.04}$\\
3.890 & 24 & 8 & 7.5 & 107.0 & $16.69^{+6.90}_{-6.27}$ && 4.750 & 23 & 13 & 11.4 & 143.3 & $4.96^{+3.66}_{-3.19}$\\
3.910 & 28 & 8 & 7.6 & 107.8 & $20.34^{+7.28}_{-6.52}$ && 4.770 & 14 & 9 & 11.5 & 144.3 & $2.37^{+2.86}_{-2.37}$\\
3.930 & 27 & 8 & 7.8 & 108.5 & $18.87^{+6.97}_{-6.27}$ && 4.790 & 15 & 12 & 11.5 & 145.3 & $2.00^{+2.36}_{-2.00}$\\
3.950 & 19 & 7 & 7.9 & 109.3 & $10.99^{+6.33}_{-4.57}$ && 4.810 & 13 & 10 & 11.6 & 146.3 & $1.32^{+2.77}_{-1.32}$\\
3.970 & 35 & 10 & 8.0 & 110.0 & $23.70^{+7.73}_{-6.92}$ && 4.830 & 26 & 13 & 11.6 & 147.3 & $6.36^{+3.51}_{-3.32}$\\
3.990 & 27 & 9 & 8.2 & 110.8 & $16.69^{+6.56}_{-5.95}$ && 4.850 & 18 & 13 & 11.7 & 148.3 & $2.09^{+3.31}_{-2.09}$\\
4.010 & 32 & 8 & 8.3 & 111.5 & $21.48^{+7.28}_{-5.88}$ && 4.870 & 17 & 8 & 11.8 & 149.3 & $4.30^{+2.70}_{-2.54}$\\
4.030 & 45 & 9 & 8.4 & 112.3 & $31.99^{+8.37}_{-7.30}$ && 4.890 & 19 & 10 & 11.8 & 150.3 & $4.25^{+2.87}_{-2.74}$\\
4.050 & 28 & 10 & 8.5 & 113.1 & $15.09^{+6.90}_{-5.19}$ && 4.910 & 15 & 10 & 11.9 & 151.3 & $2.02^{+2.90}_{-2.02}$\\
4.070 & 29 & 10 & 8.6 & 113.9 & $15.93^{+6.59}_{-5.45}$ && 4.930 & 21 & 9 & 11.9 & 152.4 & $5.23^{+3.20}_{-2.39}$\\
4.090 & 24 & 10 & 8.7 & 114.6 & $11.16^{+6.19}_{-4.68}$ && 4.950 & 19 & 11 & 12.0 & 153.4 & $3.65^{+2.81}_{-2.71}$\\
4.110 & 34 & 8 & 8.8 & 115.4 & $20.80^{+7.08}_{-5.22}$ && 4.970 & 19 & 10 & 12.0 & 154.5 & $3.76^{+3.04}_{-2.31}$\\
4.130 & 29 & 11 & 9.0 & 116.2 & $14.50^{+5.99}_{-5.50}$ && 4.990 & 20 & 13 & 12.1 & 155.5 & $2.97^{+3.02}_{-2.66}$\\
4.150 & 22 & 14 & 9.1 & 117.0 & $6.33^{+5.34}_{-5.22}$ && 5.010 & 17 & 12 & 12.1 & 156.6 & $1.91^{+2.93}_{-1.91}$\\
4.170 & 29 & 11 & 9.2 & 117.8 & $13.73^{+6.04}_{-5.05}$ && 5.030 & 14 & 11 & 12.2 & 157.7 & $1.31^{+2.40}_{-1.31}$\\
4.190 & 46 & 9 & 9.3 & 118.6 & $27.74^{+7.80}_{-5.84}$ && 5.050 & 17 & 11 & 12.2 & 158.8 & $2.30^{+2.83}_{-2.20}$\\
4.210 & 69 & 7 & 9.4 & 119.4 & $46.04^{+9.88}_{-7.45}$ && 5.070 & 20 & 13 & 12.3 & 159.8 & $2.98^{+2.74}_{-2.69}$\\
4.230 & 111 & 8 & 9.4 & 120.2 & $75.77^{+13.44}_{-10.79}$ && 5.090 & 23 & 12 & 12.4 & 161.0 & $5.06^{+2.42}_{-3.13}$\\
4.250 & 98 & 13 & 9.5 & 121.0 & $61.46^{+11.93}_{-9.67}$ && 5.110 & 19 & 14 & 12.4 & 162.1 & $1.81^{+2.94}_{-1.81}$\\
4.270 & 90 & 12 & 9.6 & 121.9 & $55.47^{+11.03}_{-8.93}$ && 5.130 & 17 & 11 & 12.5 & 163.2 & $2.47^{+2.43}_{-2.38}$\\
4.290 & 88 & 15 & 9.7 & 122.7 & $50.84^{+10.89}_{-8.45}$ && 5.150 & 15 & 12 & 12.5 & 164.3 & $0.95^{+2.61}_{-0.95}$\\
4.310 & 68 & 12 & 9.8 & 123.5 & $38.51^{+8.78}_{-7.14}$ && 5.170 & 12 & 15 & 12.6 & 165.4 & $0.40^{+0.67}_{-0.40}$\\
4.330 & 68 & 13 & 9.9 & 124.4 & $37.24^{+8.64}_{-7.05}$ && 5.190 & 25 & 12 & 12.6 & 166.6 & $5.06^{+2.92}_{-2.50}$\\
4.350 & 72 & 10 & 10.0 & 125.2 & $41.15^{+9.08}_{-6.92}$ && 5.210 & 11 & 11 & 12.7 & 167.8 & $0.39^{+1.62}_{-0.39}$\\
4.370 & 45 & 14 & 10.1 & 126.0 & $20.49^{+6.30}_{-5.65}$ && 5.230 & 28 & 18 & 12.7 & 168.9 & $3.77^{+3.15}_{-2.82}$\\
4.390 & 50 & 12 & 10.1 & 126.9 & $24.53^{+6.80}_{-5.57}$ && 5.250 & 20 & 15 & 12.8 & 170.1 & $1.93^{+2.55}_{-1.93}$\\
4.410 & 29 & 15 & 10.2 & 127.8 & $10.69^{+3.20}_{-6.38}$ && 5.270 & 14 & 11 & 12.8 & 171.3 & $1.02^{+2.23}_{-1.02}$\\
4.430 & 38 & 15 & 10.3 & 128.6 & $14.13^{+5.94}_{-4.65}$ && 5.290 & 25 & 16 & 12.9 & 172.5 & $3.65^{+2.49}_{-2.93}$\\
4.450 & 25 & 7 & 10.4 & 129.5 & $11.22^{+4.18}_{-3.74}$ && 5.310 & 19 & 17 & 12.9 & 173.7 & $1.00^{+2.24}_{-1.00}$\\
4.470 & 25 & 9 & 10.5 & 130.4 & $9.84^{+4.19}_{-3.83}$ && 5.330 & 23 & 11 & 13.0 & 174.9 & $4.32^{+2.59}_{-2.20}$\\
4.490 & 28 & 9 & 10.5 & 131.3 & $11.12^{+4.77}_{-3.55}$ && 5.350 & 14 & 12 & 13.0 & 176.1 & $1.83^{+0.95}_{-1.83}$\\
4.510 & 33 & 12 & 10.6 & 132.2 & $12.16^{+5.18}_{-3.97}$ && 5.370 & 15 & 14 & 13.0 & 177.4 & $0.12^{+2.38}_{-0.12}$\\
4.530 & 29 & 11 & 10.7 & 133.0 & $10.23^{+4.78}_{-3.64}$ && 5.390 & 18 & 16 & 13.1 & 178.6 & $0.48^{+2.55}_{-0.48}$\\
4.550 & 25 & 9 & 10.7 & 134.0 & $9.32^{+3.96}_{-3.63}$ && 5.410 & 16 & 14 & 13.1 & 179.9 & $0.71^{+2.14}_{-0.71}$\\
4.570 & 33 & 11 & 10.8 & 134.9 & $12.45^{+4.76}_{-3.95}$ && 5.430 & 16 & 11 & 13.2 & 181.1 & $1.75^{+2.02}_{-1.75}$\\
4.590 & 29 & 12 & 10.9 & 135.8 & $9.45^{+4.44}_{-3.75}$ && 5.450 & 15 & 12 & 13.2 & 182.4 & $1.04^{+1.99}_{-1.04}$\\
4.610 & 24 & 11 & 11.0 & 136.7 & $7.28^{+3.81}_{-3.57}$ && 5.470 & 18 & 15 & 13.3 & 183.7 & $0.92^{+2.30}_{-0.92}$\\
4.630 & 22 & 9 & 11.0 & 137.6 & $6.82^{+3.91}_{-2.91}$ && 5.490 & 15 & 13 & 13.3 & 185.0 & $0.68^{+1.98}_{-0.68}$\\
4.650 & 24 & 8 & 11.1 & 138.6 & $8.73^{+3.61}_{-3.28}$ && & & & & & \\
        \hline\hline
    \end{tabular}
    \label{tab:pi_comb_xsec}
\end{table*}

\begin{table*}[htbp]
    \centering
    \caption{Combined cross sections $\sigma$ of $\EE\to\kkjpsi$ of Belle~\cite{Shen:2014} and Belle~II measurements. We also list the $\EE$ c.m.\ energy ($\ecms$), number of observed events ($n^{\rm obs}$), number of backgrounds estimated from the normalized $\jpsi$ mass sidebands ($n^{\rm bkg}$), detection efficiency ($\varepsilon$), and effective ISR luminosity ($\lum$). All values are calculated for a 50 MeV bin size and $\ecms$ is the central value of the bin. The uncertainties include both statistical and systematic contributions.}
    \begin{tabular}{lcccccccccccc}
        \hline\hline
        $\ecms$ (GeV) & $n^{\rm obs}$ & $n^{\rm bkg}$ & $\varepsilon~(\%)$ & $\lum~({\rm pb}^{-1})$ & $\sigma~(\rm pb)$ && $\ecms$ (GeV) & $n^{\rm obs}$ & $n^{\rm bkg}$ & $\varepsilon~(\%)$ & $\lum~({\rm pb}^{-1})$ & $\sigma~(\rm pb)$ \\
        \hline
4.175 & 1 & 0 & 1.7 & 297.5 & $1.67^{+2.31}_{-1.17}$ && 5.125 & 15 & 2 & 11.3 & 410.9 & $2.34^{+0.88}_{-0.75}$\\
4.225 & 3 & 1 & 2.8 & 302.0 & $1.99^{+2.30}_{-1.99}$ && 5.175 & 8 & 5 & 11.5 & 418.1 & $0.52^{+0.69}_{-0.52}$\\
4.275 & 3 & 2 & 3.8 & 307.6 & $1.44^{+1.07}_{-1.44}$ && 5.225 & 12 & 4 & 11.8 & 425.3 & $1.34^{+0.76}_{-0.70}$\\
4.325 & 0 & 1 & 4.7 & 313.2 & $0.00^{+0.29}_{-0.11}$ && 5.275 & 13 & 7 & 12.0 & 432.6 & $0.97^{+0.81}_{-0.78}$\\
4.375 & 2 & 0 & 5.5 & 317.8 & $0.97^{+0.88}_{-0.54}$ && 5.325 & 6 & 8 & 12.3 & 439.9 & $0.00^{+0.31}_{-0.06}$\\
4.425 & 6 & 1 & 6.2 & 323.5 & $2.10^{+1.30}_{-1.08}$ && 5.375 & 18 & 2 & 12.6 & 447.3 & $2.39^{+0.81}_{-0.68}$\\
4.475 & 21 & 2 & 6.8 & 329.2 & $7.12^{+2.20}_{-1.86}$ && 5.425 & 8 & 8 & 12.8 & 455.7 & $0.00^{+0.63}_{-0.13}$\\
4.525 & 11 & 5 & 7.4 & 334.9 & $2.04^{+1.53}_{-1.44}$ && 5.475 & 4 & 4 & 13.0 & 464.1 & $0.00^{+0.42}_{-0.08}$\\
4.575 & 5 & 5 & 7.9 & 340.6 & $0.31^{+0.76}_{-0.31}$ && 5.525 & 13 & 4 & 13.3 & 471.7 & $1.20^{+0.63}_{-0.57}$\\
4.625 & 9 & 3 & 8.3 & 346.4 & $2.04^{+0.86}_{-1.33}$ && 5.575 & 8 & 6 & 13.6 & 480.2 & $0.26^{+0.53}_{-0.26}$\\
4.675 & 8 & 2 & 8.7 & 352.2 & $1.64^{+0.99}_{-0.87}$ && 5.625 & 5 & 7 & 13.8 & 488.9 & $0.12^{+0.09}_{-0.12}$\\
4.725 & 4 & 5 & 9.1 & 358.0 & $0.00^{+0.57}_{-0.11}$ && 5.675 & 5 & 2 & 14.1 & 497.6 & $0.36^{+0.36}_{-0.33}$\\
4.775 & 23 & 1 & 9.4 & 363.9 & $5.38^{+1.50}_{-1.24}$ && 5.725 & 5 & 3 & 14.3 & 507.3 & $0.35^{+0.25}_{-0.35}$\\
4.825 & 5 & 1 & 9.7 & 370.7 & $0.93^{+0.66}_{-0.56}$ && 5.775 & 3 & 5 & 14.5 & 516.2 & $0.00^{+0.11}_{-0.02}$\\
4.875 & 6 & 2 & 10.0 & 376.7 & $0.89^{+0.71}_{-0.64}$ && 5.825 & 5 & 5 & 14.8 & 526.1 & $0.00^{+0.37}_{-0.07}$\\
4.925 & 11 & 6 & 10.3 & 383.7 & $1.06^{+0.98}_{-0.94}$ && 5.875 & 7 & 8 & 15.0 & 536.0 & $0.00^{+0.33}_{-0.07}$\\
4.975 & 15 & 3 & 10.6 & 389.7 & $2.44^{+1.01}_{-0.88}$ && 5.925 & 6 & 6 & 15.2 & 546.1 & $0.10^{+0.28}_{-0.10}$\\
5.025 & 7 & 4 & 10.8 & 396.7 & $0.59^{+0.72}_{-0.59}$ && 5.975 & 3 & 4 & 15.4 & 556.2 & $0.00^{+0.18}_{-0.04}$\\
5.075 & 12 & 4 & 11.1 & 403.8 & $1.69^{+0.67}_{-0.97}$ && & & & & & \\
        \hline\hline
    \end{tabular}
    \label{tab:k_comb_xsec}
\end{table*}

\section*{Acknowledgements}
% % ------------------------------
% %    Founding and Acknowledge
% % ------------------------------
% Policy from October 20, 2022
This work, based on data collected using the Belle~II detector, which was built and commissioned prior to March 2019,
%Belle1 and data collected using the Belle detector, which was operated until June 2010,
was supported by
%Armenia
Higher Education and Science Committee of the Republic of Armenia Grant No.~23LCG-1C011;
%Australia
Australian Research Council and Research Grants
No.~DP200101792, % Jackson
No.~DP210101900, % Urquijo
No.~DP210102831, % Sevior
No.~DE220100462, % Hsu
No.~LE210100098, % Infrastructure
and
No.~LE230100085; % Infrastructure
%Austria
Austrian Federal Ministry of Education, Science and Research,
Austrian Science Fund (FWF) Grants
DOI:~10.55776/P34529,
DOI:~10.55776/J4731,
DOI:~10.55776/J4625,
DOI:~10.55776/M3153,
and
DOI:~10.55776/PAT1836324,
and
Horizon 2020 ERC Starting Grant No.~947006 ``InterLeptons'';
%Canada
Natural Sciences and Engineering Research Council of Canada, Digital Research Alliance of Canada, and Canada Foundation for Innovation;
%China
National Key R\&D Program of China under Contract No.~2024YFA1610503,
and
No.~2024YFA1610504
National Natural Science Foundation of China and Research Grants
No.~11575017,
No.~11761141009,
No.~11705209,
No.~11975076,
No.~12135005,
No.~12150004,
No.~12161141008,
No.~12335004, % Yuan, IHEP
No.~12405099,
No.~12475093,
and
No.~12175041,
and Shandong Provincial Natural Science Foundation Project~ZR2022JQ02;
%Czech Republic
the Czech Science Foundation Grant No. 22-18469S,  Regional funds of EU/MEYS: OPJAK
FORTE CZ.02.01.01/00/22\_008/0004632 
and
Charles University Grant Agency project No. 246122;
%EU
European Research Council, Seventh Framework PIEF-GA-2013-622527,
Horizon 2020 ERC-Advanced Grants No.~267104 and No.~884719,
Horizon 2020 ERC-Consolidator Grant No.~819127,
Horizon 2020 Marie Sklodowska-Curie Grant Agreement No.~700525 ``NIOBE''
and
No.~101026516,
and
Horizon 2020 Marie Sklodowska-Curie RISE project JENNIFER2 Grant Agreement No.~822070 (European grants);
%France
L'Institut National de Physique Nucl\'{e}aire et de Physique des Particules (IN2P3) du CNRS
and
L'Agence Nationale de la Recherche (ANR) under Grant No.~ANR-23-CE31-0018 (France);
%Germany
BMFTR, DFG, HGF, MPG, and AvH Foundation (Germany);
%India
Department of Atomic Energy under Project Identification No.~RTI 4002,
Department of Science and Technology,
and
UPES SEED funding programs
No.~UPES/R\&D-SEED-INFRA/17052023/01 and
No.~UPES/R\&D-SOE/20062022/06 (India);
%Israel
Israel Science Foundation Grant No.~2476/17,
U.S.-Israel Binational Science Foundation Grant No.~2016113, and
Israel Ministry of Science Grant No.~3-16543;
%Italy
Istituto Nazionale di Fisica Nucleare and the Research Grants BELLE2,
and
the ICSC – Centro Nazionale di Ricerca in High Performance Computing, Big Data and Quantum Computing, funded by European Union – NextGenerationEU;
%Japan
Japan Society for the Promotion of Science, Grant-in-Aid for Scientific Research Grants
No.~16H03968,
No.~16H03993,
No.~16H06492,
No.~16K05323,
No.~17H01133,
No.~17H05405,
No.~18K03621,
No.~18H03710,
No.~18H05226,
No.~19H00682, % Niigata
No.~20H05850,
No.~20H05858,
No.~22H00144,
No.~22K14056,
No.~22K21347,
No.~23H05433,
No.~26220706,
and
No.~26400255,
%the National Institute of Informatics, and Science Information NETwork 5 (SINET5), 
and
the Ministry of Education, Culture, Sports, Science, and Technology (MEXT) of Japan;  
%Korea
National Research Foundation (NRF) of Korea Grants
No.~2021R1-F1A-1064008, 
No.~2022R1-A2C-1003993,
No.~2022R1-A2C-1092335,
No.~RS-2016-NR017151,
No.~RS-2018-NR031074,
No.~RS-2021-NR060129,
No.~RS-2023-00208693,
No.~RS-2024-00354342
and
No.~RS-2025-02219521,
Radiation Science Research Institute,
Foreign Large-Size Research Facility Application Supporting project,
the Global Science Experimental Data Hub Center, the Korea Institute of Science and
Technology Information (K25L2M2C3 ) 
and
KREONET/GLORIAD;
%Malaysia
Universiti Malaya RU grant, Akademi Sains Malaysia, and Ministry of Education Malaysia;
%Mexico
% CINVESTAV-IPN, UNAM, UAS, BUAP and CONACYT are funded under
Frontiers of Science Program Contracts
No.~FOINS-296,
No.~CB-221329,
No.~CB-236394,
No.~CB-254409,
and
No.~CB-180023, and SEP-CINVESTAV Research Grant No.~237 (Mexico);
%Poland
the Polish Ministry of Science and Higher Education and the National Science Center;
%Russia
the Ministry of Science and Higher Education of the Russian Federation
and
the HSE University Basic Research Program, Moscow;
%Saudi Arabia
University of Tabuk Research Grants
No.~S-0256-1438 and No.~S-0280-1439 (Saudi Arabia), and
Researchers Supporting Project number (RSPD2025R873), King Saud University, Riyadh,
Saudi Arabia;
%Slovenia
Slovenian Research Agency and Research Grants
No.~J1-50010
and
No.~P1-0135;
%Spain
Ikerbasque, Basque Foundation for Science,
State Agency for Research of the Spanish Ministry of Science and Innovation through Grant No. PID2022-136510NB-C33, Spain,
Agencia Estatal de Investigacion, Spain
Grant No.~RYC2020-029875-I
and
Generalitat Valenciana, Spain
Grant No.~CIDEGENT/2018/020;
%Swiss (Belle 1)
%Belle1 the Swiss National Science Foundation;
%Sweden
The Knut and Alice Wallenberg Foundation (Sweden), Contracts No.~2021.0174, No.~2021.0299, and No.~2023.0315;
%Taiwan
National Science and Technology Council,
and
Ministry of Education (Taiwan);
%Thailand
Thailand Center of Excellence in Physics;
%Turkey
TUBITAK ULAKBIM (Turkey);
%Ukraine
National Research Foundation of Ukraine, Project No.~2020.02/0257,
and
Ministry of Education and Science of Ukraine;
%USA
the U.S. National Science Foundation and Research Grants
No.~PHY-1913789 % Indiana CEEM
and
No.~PHY-2111604, % Luther
and the U.S. Department of Energy and Research Awards
No.~DE-AC06-76RLO1830, % PNNL
No.~DE-SC0007983, % Wayne State
No.~DE-SC0009824, % Florida
No.~DE-SC0009973, % VPI
No.~DE-SC0010007, % Duke
No.~DE-SC0010073, % South Carolina
No.~DE-SC0010118, % Carnegie Mellon
No.~DE-SC0010504, % Hawaii
No.~DE-SC0011784, % Cincinnati
No.~DE-SC0012704, % BNL
No.~DE-SC0019230, % Duke
No.~DE-SC0021274, % Mississippi
No.~DE-SC0021616, % Mississippi
No.~DE-SC0022350, % Louisville
No.~DE-SC0023470; % South Alabama
%last group
and
%Vietnam
the Vietnam Academy of Science and Technology (VAST) under Grants
No.~NVCC.05.02/25-25
and
No.~DL0000.05/26-27.

% Policy from October 20, 2022
These acknowledgements are not to be interpreted as an endorsement of any statement made
by any of our institutes, funding agencies, governments, or their representatives.

We thank the SuperKEKB team for delivering high-luminosity collisions;
the KEK cryogenics group for the efficient operation of the detector solenoid magnet and IBBelle on site;
the KEK Computer Research Center for on-site computing support; the NII for SINET6 network support;
and the raw-data centers hosted by BNL, DESY, GridKa, IN2P3, INFN, 
%Belle1 PNNL/EMSL, 
and the University of Victoria.

% -----------------------
%       References
% -----------------------
\renewcommand{\baselinestretch}{1.2}

%%%%%%%%%%%%%%%%%%%%%%%%%%%%%%%%%%%%%%%%%%%%%%%%%%%%%%%%%%%%%%%%%%%%%%%%%%%%
\end{document}